\documentclass[aps,pra,twocolumn,superscriptaddress,floatfix,tightenlines,showpacs,notitlepage]{revtex4-2}

\usepackage[utf8]{inputenc}
\usepackage[american,british]{babel}
\usepackage[T1]{fontenc}
\usepackage[pdftex]{graphicx}  
\usepackage{xcolor}
\usepackage{dcolumn}
\usepackage{physics}
\usepackage{braket}
\usepackage{bm}
\usepackage{amsmath,amsthm,amssymb}
\usepackage{color}
\usepackage{verbatim}
\usepackage[normalem]{ulem}
\usepackage{dsfont}

\usepackage{soul}
\usepackage{hyperref}
\hypersetup{
 colorlinks=true,
 linkcolor=blue,
 anchorcolor = blue,
 citecolor = blue,
 filecolor = blue,
 urlcolor = blue
}

\begin{document}

\title{
Towers of quantum many-body scars under stochastic resetting}

\author{Lorenzo Gotta}
\affiliation{Department of Quantum Matter Physics, University of Geneva, 1211 Geneva, Switzerland.}

\author{Manas Kulkarni}
\affiliation{International Centre for Theoretical Sciences, Tata Institute of Fundamental
Research, Bangalore 560089, India.}

\author{Gabriele Perfetto}
\affiliation{Institut für Theoretische Physik, ETH Zürich, Wolfgang-Pauli-Str. 27, 8093 Zürich, Switzerland.}

\begin{abstract}
Towers of quantum many-body scars are sets of highly-excited eigenstates of nonintegrable Hamiltonians whose dynamics shows athermal behavior and persistent oscillations in time. The preparation of such states is, however, challenging due to their entanglement content.  
In this work, we show that local properties of such states can be prepared by interspersing the scarred dynamics with stochastic resets to much simpler unentangled product states. Stochastic resetting amounts to reinitializing the many-body wavefunction of the system at random times to a predefined state, which we choose to be in the scarred subspace. We derive several analytical results for the ensuing dynamics, e.g., for the time evolution of the fidelity and of local observables. Resetting damps the scarred oscillations and generates spatial off-diagonal long-range order in the ensuing stationary state. The latter shows mixedness that scales logarithmically as a function of the system size, which follows from the structure of the scarred eigenstates. We prove that such stationary states are locally equivalent, in the sparse-resetting limit, to a single pure scarred eigenstate, which is determined by the reset state. This protocol thereby might represent a route to the experimental preparation of the local properties of correlated and entangled states through resetting.
\end{abstract}
\maketitle 
\section{Introduction}
Upon time-evolution generated by a quantum many-body Hamiltonian, any information stored in the initial state is typically washed out and the system becomes locally equivalent to a thermal state, described by the proper Gibbs density matrix \cite{Rigol_2008,Mori_2018}. The pillar of the understanding of such thermalization process is the eigenstate thermalization hypothesis (ETH) \cite{Dalessio_2016, Deutsch_2018}. According to the latter, eigenstates within the same energy shell are locally equivalent.     
This paradigm is, however, violated, e.g., by the so-called scarred eigenstates \cite{Serbyn_2021,Moudgalya_2022,Chandran_2023}. The latter are exceptional energy eigenstates in the middle of the spectrum. Their atypical behavior, compared to the ETH-satisfying eigenstates, lies, for instance,
in their slower entanglement scaling than volume law, and in the presence of long-range correlations \cite{Choi_2019,Iadecola_2019,Schecter_2019,Mark_2020_2,Iadecola_2020,Gotta_2023,Sanada_2023,Imai_2025}. These are, indeed, necessarily absent in finite temperature eigenstates obeying ETH \cite{Yang_1962, Yang_1989,Bruno_2013,Watanabe_2015}. Recently, these phenomena 
have been observed in experiments with quantum quenches of Rydberg atoms \cite{Bernien_2017,Bluvstein_2021,Su_2023,Harris_2025}. These experiments are described by the PXP model \cite{Turner_2018,Turner_2018_2}, which cannot be analytically tackled.  
Analytically tractable settings of the scarred physics is, on the other hand, possible 
in terms of towers of equally-spaced eigenstates obtained by adding quasiparticle excitations on top of a low-entangled reference state \cite{EE_scars_FH,Iadecola_2019,Schecter_2019,Iadecola_2020,Shibata_2020,Mark_2020,Gotta_2022}. The algebraic structure of such states allows, as a matter of fact, to analytically determine, for instance, the logarithmic growth of their entanglement \cite{Odea_2020}. 

Even in this simplified setting, it remains, however, 
challenging to harness and prepare such eigenstates due to their correlated nature~\cite{Buca_2019,Wang_2024,Marche_2025,Garcia_2025,Gotta_2025}. A largely unexplored opportunity, in this direction, is offered by stochastic resetting protocols \cite{EM2011,EM2011a,Evans_2020,Gupta_2022,Nagar_2023,Kundu_2024}. Stochastic resetting has been widely investigated in the context of classical stochastic processes, while in quantum systems it received attention only recently~\cite{Mukherjee_2018,Rose2018spectral}. In particular, resetting in quantum systems has been shown to be beneficial to expedite the first-hitting time of quantum walks \cite{grunbaum2013recurrence,Barkai2017,Barkai2018,tornow2023,majumdar2023,Walter25,yin2025restart,restart_Yin,Modak_resetting,roy2025causality,king2026time} and to engineer correlated stationary states \cite{Hartmann2006,Linden2010,Rose2018spectral,Mukherjee_2018,Perfetto_2021,Magoni_2022,Perfetto_2022,dattagupta2022stochastic,Anish_conditional,Kulkarni_2023,kulkarni2025dynamically,wald2025stochastic,soldner2025nonanaliticities,kulkarni2025dynamically,bao_accelerating,solanki2025universal}. The effect of stochastic resetting on quantum dynamics, as a matter of fact, amounts to monitoring the system, which drives the dynamics to a non-equilibrium stationary state (NESS). The latter can be expressed as a non-diagonal ensemble in the energy eigenstates \cite{Mukherjee_2018,Perfetto_2021,Kulkarni_2023}. It is therefore natural to investigate the interplay between stochastic resetting and quantum many-body scarred dynamics. In particular, the possible connection between the NESS built from such atypical eigenstates and the dynamical preparation and stabilization of scarred eigenstates is not understood.    

In this work, we address these points. We do this by considering the analytically tractable case of Hamiltonians hosting towers of equally-spaced eigenstates. These states are generated by the repeated application of a suitable quasiparticle creation operator, which obeys together with the Hamiltonian a restricted spectrum-generating algebra \cite{Mark_2020,Odea_2020,Chandran_2023}. Concrete examples are given by the XY spin-$1$ chain and deformed Fermi-Hubbard models.  We superimpose stochastic resets to such Hamiltonian scarred dynamics. One considers, for simplicity, coherent-like reset states \cite{Shibata_2020,Liska_2023}, which are identified by a complex parameter. These states are entirely contained within the scarred subspace and are uncorrelated in real space. We obtain exact results on the behavior of dynamical proxies, such as the time evolution of fidelities and local observables. Both cases showcase that stochastic resetting drives the system to a NESS, thereby damping the periodic revivals of the scarred quench dynamics. Resetting thus leads to a damping, and eventual loss, of the time-crystalline order manifested by dynamical correlations functions in scarred eigenstates. On the other hand, we surprisingly show that resetting generates spatial off-diagonal long-range order and condensation of quasi-particles in the NESS. The NESS is a statistical mixture of scarred eigenstates. We quantify this mixedness by analytically computing the Rényi entropies of the whole NESS density matrix. For such entropies we find a logarithmic scaling as a function of the system size. We interpret such a result by showing that the NESS density matrix has the same structure as that of the reduced density matrix of a scarred eigenstate. The latter gives an atypically low, logarithmic in system size, entanglement. The NESS generated therefore directly inherits the atypically low entanglement content of scarred eigenstates in terms of an
atypically low mixedness.
We then show that in the sparse-resetting the NESS reduces to a diagonal ensemble over the scarred states. In the large volume limit, this ensemble becomes \textit{locally} equivalent to an individual scarred eigenstate. This means that averages of local observables over the NESS coincide with those obtained by averaging over a single pure scarred eigenstate, which is uniquely identified by the complex parameter labeling the coherent-like reset state.  
This local equivalence between diagonal ensemble in the scar subspace and expectation values over a single scarred eigenstate is in agreement with the theoretical result in Ref.~\cite{Morettini_2025}, albeit we can here dynamically generate it through a dissipative resetting protocol. 

Our work thereby fits within the framework of nonequilibrium protocols aimed at preparing and manipulating correlated quantum many-body scarred eigenstates; this problem has recently received attention, e.g., in Ref.~\cite{Yin_2025}, where tunable superpositions of the scarred eigenstates are globally produced by periodically interrupting the unitary evolution with a global post-selected projective measurements. In our protocol, scarred eigenstates are, differently, synthesized only locally. This local state preparation protocol crucially does not require postselection of measurement outcomes, which could result into severe overheads. On the contrary, only few and rare stochastic events are here needed. Moreover, reset projections are performed to uncorrelated product states, which are therefore experimentally realizable. The present resetting protocol to uncorrelated coherent-like states might thus offer a viable method to engineer and locally prepare correlated quantum many-body scar eigenstates. 

The rest of the manuscript is organized as follows. In Sec.~\ref{sec:model}, we introduce the models  that we investigate in the manuscript: the deformed Fermi-Hubbard and the spin-$1$ $XY$ chain. Both these models possess towers of many-body quantum scars. In Sec.~\ref{sec:3resetting}, we introduce the resetting protocol to a coherent-like state entirely contained in the scar manifold. In Sec.~\ref{sec:fidelity}, we give results for the dynamics of the fidelity between the reset state and the time-evolved state. In Sec.~\ref{sec:correlations}, we discuss the effect of resetting on mean values of local observables and correlation functions. In Sec.~\ref{sec:entropy_mix}, we compute the dynamics and the stationary Rényi entropies in the stationary state induced by resetting. In Sec.~\ref{sec:6equivalence}, we show that in the weak resetting limit, such stationary state is locally indistinguishable from a single scar eigenstate. In Sec.~\ref{sec:conclusions}, we eventually draw our conclusions. Details about the calculations are consigned to the various Appendix sections. 

\section{Model}
\label{sec:model}
Recent progress in the characterization of ETH violation and ergodicity breaking in quantum non-integrable many-body Hamiltonians led to the discovery of quantum many-body scars. These are atypical highly excited eigenstates embedded within the energy spectrum of ETH-satisfying ones after resolving all symmetries of the underlying Hamiltonian, see, e.g., Refs.~\cite{Moudgalya_2022,Chandran_2023} for reviews on the subject. In this manuscript, we will specifically consider towers of many-body scars that are related to the presence of stable quasi-particles. The latter are created by an operator $\hat Q^{\dagger}$ acting on a reference vacuum state $\ket{\psi_0}$. The quasi-particle creation operator $\hat Q^{\dagger}$ obeys together with the Hamiltonian $\hat H$ the so-called restricted spectrum generating algebra:
\begin{equation}
\label{Eq:sga}
    [\hat H, \hat Q^{\dag}]\ket{\psi_n} = -\omega \hat Q^{\dag}\ket{\psi_n},
\end{equation}
for some $\omega\in\mathbb{R}$ and for values of $n=0,1,\dots N$ such that 
\begin{equation}
\ket{\psi_n}\propto (\hat Q^{\dag})^n \ket{\psi_0}\neq 0, \quad (\hat Q^{\dag})^{N+1}\ket{\psi_0}=0.
\label{eq:scars_towers}
\end{equation}
Under the further assumption that $\ket{\psi_0}$ is an eigenstate of $\hat H$ with eigenvalue $E_0$, Eq.\eqref{Eq:sga} implies that $\ket{\psi_n}$, namely the state with $n$ quasiparticles on top of the vacuum state $\ket{\psi_0}$, is an eigenstate of $\hat H$ with eigenvalue $E_0-n\omega$. The energy scale $\omega$ is therefore the constant energy spacing between consecutive quantum scars. For the ladder operators $\hat Q^{\dag}$ we will consider the form
\begin{equation}
\hat Q^{\dag} = \sum_j e^{ikj} \hat q^{\dag}_j,
\label{eq:ladder_quasiparticle}
\end{equation}
with $(\hat q^{\dag}_j)^2 =0$ and $[\hat q^{\dag}_j, \hat q^{\dag}_l]=0,\,\forall j,l$. For the Hamiltonian, we will consider two paradigmatic examples. 

First, the Fermi-Hubbard model with correlated pair hopping~\cite{Mark_2020_2}:
\begin{align}
\label{Eq:F_H_model}
\hat H =& -\sum_{j,\sigma}\left[t-t'(\hat n_{j,-\sigma} + \hat n_{j+1,-\sigma})\right](\hat c^{\dag}_{j,\sigma}\hat c_{j+1,\sigma}+\mathrm{h.c.}) \nonumber \\
&+U\sum_j \hat n_{j,\uparrow} \hat n_{j,\downarrow}-\mu\sum_{j,\sigma}\hat n_{j,\sigma},
\end{align}
where $j=1,2,\dots L$, labels the lattice site in one dimension, and periodic boundary conditions. Here, $t$ is the hopping amplitude of the fermionic particles annihilated (created) by the operator $\hat c_{j,\sigma}$ ($\hat c_{j,\sigma}^{\dagger}$) with spin $\sigma=\{\uparrow,\downarrow \}$ at lattice site $j$.   The coupling constant $U$ is the onsite interaction energy between fermions with opposite spins, while $\mu$ is the chemical potential. In the absence of the correlated hopping $t'=0$, the model is integrable and it displays on bipartite lattices an $SU(2) \times SU(2)$ symmetry associated to spin and pseudospin symmetry \cite{Yang_1989,SO4_hubbard}. The pseudospin symmetry, namely, allows to identify the so-called eta-pairing eigenstates $\ket{\eta_n}$, 
\begin{equation}
\ket{\eta_n}=C_n \,(\hat \eta^{\dagger})^n \ket{0}, \quad n=0,1,\dots L,
\end{equation}
with the normalization factor $C_n=\sqrt{(L-n)! /(n! \, L!)}$. The quasi-particle creation operation
\begin{equation}
\hat \eta^{\dagger}=\sum_j e^{i\pi j} \hat c^{\dagger}_{j,\uparrow}\hat c^{\dagger}_{j,\downarrow}, 
\end{equation}
creates a doublon (doubly-occupied site) at momentum $k=\pi$ on top of the fermionic vacuum $\ket{0}$. At $t'\neq 0$, integrability and pseudospin conservation are broken. Spin conservation and the eta-pairing states are still preserved and continue to be eigenstates of the Hamiltonian \cite{Mark_2020_2}. Namely. they have been identified as representing a tower of quantum many-body scars according to the definitions in Eqs.~\eqref{Eq:sga}-\eqref{eq:ladder_quasiparticle}, with the identification $Q^{\dagger}=\eta^\dagger$ for the ladder operator within the scar tower. The respective energy spacing is given by $\omega=2\mu - U$. These states will form the basis of our analysis for the Fermi-Hubbard model. 

Second, we consider the spin-$1$ $XY$ model~\cite{Schecter_2019} on a one-dimensional lattice $j=1,2,\dots L$ and periodic boundary conditions:
\begin{align}
\label{Eq:spin_1_XY_model}
        \hat H &= J_1\sum_j \left(\hat S_j^x\hat S_{j+1}^x +\hat S_j^y\hat S_{j+1}^y \right)+h\sum_j \hat S_j^z + \nonumber \\
        &+J_3\sum_j \left(\hat S_j^x\hat S_{j+3}^x +\hat S_j^y\hat S_{j+3}^y \right)+D\sum_j \left(\hat S_j^z \right)^2.
\end{align}
Here $S_j^{x,y,z}$ are spin-$1$ matrices at site $j$ in the $x,y,z$ component. We take as local computation basis for each lattice site $j$ the three eigenstates of $S_j^z$ with eigenvalue $0$ ($\ket{0}$) and $\pm 1$ ($\ket{\pm 1}$). For this model a tower of many-body scars $\ket{S_n}$ is obtained
\begin{equation}
\ket{S_n}=C_n  (S^+)^n\ket{\Downarrow}, \quad n=0,1,\dots L,
\label{eq:tower_xy}
\end{equation}
where the ladder operator $\hat Q^{\dagger}=\hat S^+$ creates, in this case, a bimagnon quasiparticle (double flipped) with momentum $k=\pi$  
\begin{equation}
\hat S^+= \sum_j \frac{e^{i\pi j}}{2}\left(\hat S^+_j\right)^2, 
\end{equation}
by acting on the vacuum state $\ket{\Downarrow}=\otimes_{j=1}^L\ket{-1}_j$. The energy spacing between adjacent states in the scar tower is for this model $\omega=-2h$. 

As noted in Ref.~\cite{Mark_2020_2}, an exact correspondence between towers of scars in the XY spin-1 chain and in the Fermi-Hubbard model can be drawn. In particular, the eta-pairing creation operator $\eta^{\dagger}$ in the latter corresponds to the bimagnon creation operator $\hat{S}^+$ in the former. In addition, the vacuum state $\ket{0}$ of the Fermi-Hubbard directly maps in the state $\ket{\Downarrow}$ of the spin-1 chain. We denote in the rest of the manuscript the reference vacuum state on top of which quasi-particles are created as $\ket{\emptyset}$. Our results hereafter therefore apply for both class of models once the model-specific form for the  Hamiltonian $\hat H$, the energy spacing $\omega$ between the adjacent scar states, and the quasi-particle creation operator $\hat{Q}^{\dagger}$, are chosen. In what follows, we will study the impact of stochastic resetting onto the dynamics dictated by an Hamiltonian hosting a tower of many-body quantum scars.

\section{Stochastic resetting protocol}
\label{sec:3resetting}
Stochastic resets reinitialize the many-body wavefunction to a prescribed reset state $\ket{\psi_r}$. Henceforth, we consider the case where the reset state coincides with the initial state $\ket{\psi(0)}$. We take unentangled reset states that are product states in real space and are entirely contained within the subspace spanned by the scars. Such states can be represented as coherent-like states $\ket{\alpha}$ labeled by a complex parameter $\alpha$ as follows \cite{Mark_2020_2,Iadecola_2020,Shibata_2020,deformed_symmetry}:
\begin{align}
\label{Eq:coh_state}
\ket{\psi_r}&=\ket{\psi(0)}=\ket{\alpha} \nonumber \\
&=\frac{1}{(1+|\alpha|^2)^{L/2}} \sum_{n=0}^L \frac{\alpha^n}{n!}  (\hat Q^{\dagger})^n \ket{\emptyset} \nonumber \\
&=\frac{1}{(1+|\alpha|^2)^{L/2}}e^{\alpha \hat Q^{\dag}} \ket{\emptyset}\nonumber \\
&=\frac{1}{(1+|\alpha|^2)^{L/2}} \prod_{j=1}^L (1+\alpha e^{ikj}\hat q^{\dag}_j)\ket{\emptyset},
\end{align}
The motivation behind this choice for the reset state is twofold. First, such reset states are uncorrelated and factorized in real space. This renders their preparation and the implementation of the reset protocol easier to experimentally achieve. Second, the states are entirely contained within the scarred manifold. The ensuing reset protocol thereby allows to single out the impact of scarred states to the NESS obtained via resetting.

The time $\tau$ elapsing between consecutive resets are independent and identically distributed random variables according to an exponential waiting time distribution $p_r(\tau)$ at constant rate $r$ (Poissonian resetting)
\begin{equation}
p_r(\tau)= \theta(\tau)  r  e^{-r \tau}. 
\label{eq:Poissonian}
\end{equation}
All the results of the manuscript, however, do not depend on the specific form of the waiting time distribution. Under the resetting protocol, the dynamics is not anymore unitary and the many-body state of the system is thus described by the density matrix $\hat\rho_r(t)$. The latter can be written in terms of the density matrix $\hat{\rho}(\tau)$ associated to the reset-free unitary dynamics via the \textit{last renewal equation} \cite{Mukherjee_2018,Perfetto_2021,Magoni_2022,Perfetto_2022,dattagupta2022stochastic,Kulkarni_2023,kulkarni2025dynamically}:
\begin{equation}
\label{Eq:renewal_eq}
    \hat\rho_r(t) = e^{-rt}\hat\rho(t) +r\int_{0}^t d\tau \, e^{-r\tau} \hat \rho(\tau),
\end{equation}
Here, we denoted the unitarily evolved density matrix of the system $\hat\rho(t)=e^{-i\hat H t} \ket{\psi_r}\bra{\psi_r}e^{i\hat H t}$ starting from the reset state $\ket{\psi_r}$. The physical meaning of Eq.~\eqref{Eq:renewal_eq} is transparent: the first term takes into account the contribution of trajectories for which no reset event occurred, which have probability $e^{-rt}$, while the second term is a sum over all contributions for which the last reset event happened with rate $r$ at time $t-\tau$. After this event, the system follows the reset-free dynamics for a duration $\tau$. Despite the dynamics in Eq.\eqref{Eq:renewal_eq} not being solvable for a generic choice of the quantum many-body Hamiltonian and the initial state, the chosen setting will allow for several exact results concerning the stationary state. The non-equilibrium stationary state (NESS) density matrix $\hat{\rho}_r(+\infty)$ induced by resetting is obtained by taking the limit $t\rightarrow +\infty$ of Eq.~\eqref{Eq:renewal_eq} and it reads as
\begin{equation}
\label{Eq:NESS}
 \hat\rho_r(+\infty)= r\int_0^{+\infty} d\tau\, e^{-r\tau} \hat\rho(\tau).  
\end{equation}
To gain insight about the structure of the steady state presented in Eq.\eqref{Eq:NESS}, one can expand the reset state over the complete basis of eigenstates $\ket{m}$ of the Hamiltonian, i.e., $\ket{\psi_r}=\sum_m c_m \ket{m}$, with the overlaps $c_m=\braket{m|\psi_r}$; plugging this expression into Eq.\eqref{Eq:NESS}, one obtains the following representation of $\hat\rho_r (+\infty)$ in the basis of the eigenstates $\ket{n}$ of the Hamiltonian:
\begin{equation}
\label{Eq:NESS_E_basis}
    \hat\rho_r (+\infty) = \sum_{m,n} c_m c_n^* \frac{r}{r+i(E_m - E_n)}\ket{m}\bra{n}.
\end{equation}
From such expression, one notices that NESS obtained in the presence of resetting is off-diagonal in the energy eigenbasis for any finite resetting $r$. This property motives the nonequilibrium nature of the stationary state, which is, indeed, markedly different from the diagonal ensemble. The latter typically describes local relaxation under unitary dynamics and it is a mixed state $\hat{\rho}_{\mathrm{diag}}$ diagonal in the Hamiltonian eigenbasis \cite{Dalessio_2016}. It can be reobtained from \eqref{Eq:NESS_E_basis} by considering the sparse resetting limit $r\to 0^{+}$:
\begin{equation}
\lim_{r\to 0^+} \hat{\rho}_r(+\infty)=\hat\rho_{diag}(\ket{\psi_r})=\sum_n |c_n|^2 \ket{n}\bra{n}.
\label{eq:diagonal_ensemble}
\end{equation}
Note that the limit of vanishing resetting rates does not imply that resetting is absent. The expression in Eq.~\eqref{Eq:NESS} has been, indeed, obtained by taking the long-time limit first and assuming a finite resetting rate (the limits $t\to \infty$ and $r\to 0^+$ do not commute). In Eq.~\eqref{eq:diagonal_ensemble}, thus, resetting is still present, but reset events are sparse in time. This means that the typical timescale $\sim r^{-1}$ between consecutive resetting events is larger than any other time-scale in the system's dynamics. Importantly, in Eq.~\eqref{eq:diagonal_ensemble}, the notation stresses the functional dependence of the stationary state on the reset state. The latter selects, as a matter of fact, the eigenstates $\ket{n}$ appearing in the expression via the nonvanishing overlaps $c_n=\braket{n|\psi_r}$ between the Hamiltonian eigenspectrum and the reset state. The latter is chosen as in Eq.~\eqref{Eq:coh_state} and it therefore selects only scarred eigenstates. The stationary state \eqref{eq:diagonal_ensemble} is therefore a mixture of scarred eigenstates only and it allows to identify the effect of such atypical eigenstates on the stationary properties obtained in the presence of resetting. 

\begin{figure}
\centering
\includegraphics[width=1\columnwidth]{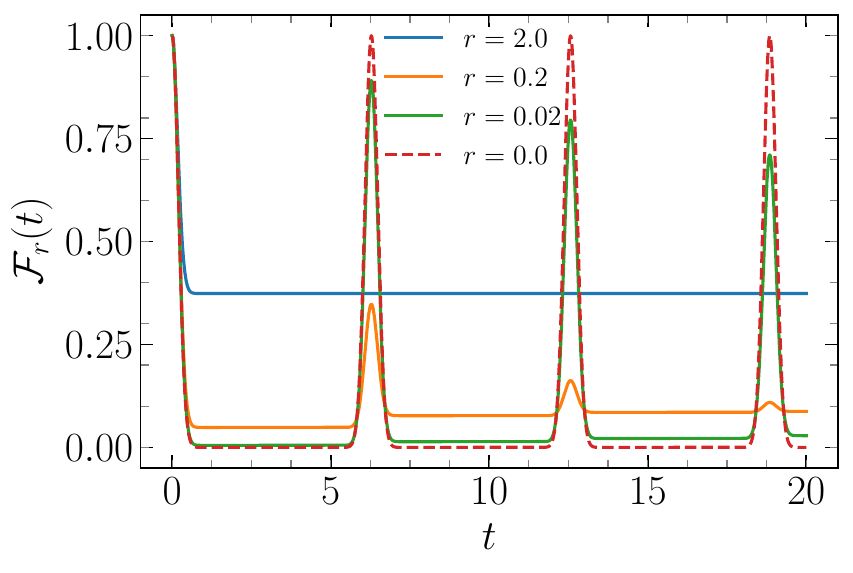}
\caption{\textbf{Fidelity between the time-evolved state and the reset state as a function of time}. 
Plot of the fidelity \eqref{eq:fidelity_def} as function of time $t$ for different values of the resetting rate: $r=0.0$ (purely unitary evolution), $r=0.02$, $r=0.2$ and $r=2.0$. In the unitary case, the fidelity shows perfects revivals of the reset state as a consequence of the equispaced scarred energy levels. For any nonzero resetting rate, the frequency of the oscillations is still visible, but the associated amplitude gets damped in time and the system eventually reaches a NESS. The other parameters are fixed as $\alpha=1.0$ for the reset state, $\omega=1.0$ and $L=50$ for the system size.}
\label{Fig:fidelity}
\end{figure}

\section{Fidelity}
\label{sec:fidelity}
An important dynamical signature of scarred eigenstates is the presence of persistent revivals at finite times and for arbitrary values of the system size $L$ in the quench dynamics ensuing from selected initial states. In order to assess the impact of stochastic resetting on such revivals, we compute the fidelity $\mathcal{F}_r(t)$ between the initial state $\ket{\alpha}$ in Eq.~\eqref{Eq:coh_state} and the time-evolved state $\hat{\rho}_r(t)$ in the presence of resetting:
\begin{equation}
    \mathcal{F}_r(t) = \bra{\alpha} \hat\rho_r(t) \ket{\alpha}.
\label{eq:fidelity_def}
\end{equation}
For purely unitary evolution $\hat{\rho}_r(t)=\hat{\rho}(t)$, according to  definition after Eq.~\eqref{Eq:renewal_eq}, and the fidelity reduces to the Loschmidt echo. Inserting the definition \eqref{eq:fidelity_def} into the renewal equation~\eqref{Eq:renewal_eq}, one obtains, 
\begin{equation}
\mathcal{F}_r(t)= e^{-rt} \mathcal{F}(t)+r\int_0^t d\tau\, e^{-r\tau} \mathcal{F}(\tau).
\label{eq:renewal_fidelity}
\end{equation}
This is a renewal equation that allows to compute the fidelity in the presence of resetting $\mathcal{F}_r(t)$ on the basis of the knowledge of the fidelity $\mathcal{F}(t)=|\braket{\alpha|\alpha(t)}|^2$ of the reset-free unitary quench dynamics. The latter is given by (we assume for simplicity and without loss of generality that $\alpha=1$) by the simple relation \cite{Schecter_2019,Mark_2020_2,Iadecola_2020}: 
\begin{equation}
\mathcal{F}(t)=|\braket{\alpha=1|\alpha=e^{i\omega t}}|^2=\cos^{2L}\left(\frac{\omega t}{2}\right),
\label{eq:fidelity_unitary_revivals}
\end{equation}
where we used that the state $\ket{\alpha}$ in Eq.~\eqref{Eq:coh_state} evolves under the unitary dynamics only through a rotation in the complex plane of the parameter $\alpha$ \cite{Shibata_2020} as 
\begin{equation}
\label{eq:coh_state_evolution}
e^{-i H t}\ket{\alpha}=\ket{e^{i \omega t} \alpha}\equiv \ket{\alpha(t)},
\end{equation}
with $\alpha(t)=\mbox{exp}(i \omega t)\alpha$. The expression in Eq.~\eqref{eq:fidelity_unitary_revivals} neatly shows the revivals, with period $T=2\pi/\omega$, ensuing from the quench dynamics from the state \eqref{Eq:coh_state}. In this case revivals are perfect and the fidelity goes periodically back to $1$ because the state $\ket{\alpha}$ is entirely contained within the scarred subspace. Upon inserting the result \eqref{eq:fidelity_unitary_revivals} into the renewal equation \eqref{eq:renewal_fidelity}, one can therefore readily compute the fidelity in the presence of resetting. The associated expression attains a stationary value $\mathcal{F}_r(+\infty)$ for long times (cf. Appendix \ref{S:fidelity} for the details of the calculations):
\begin{equation}
    \mathcal{F}_r(+\infty) = \frac{1}{2^{2L}}\sum_{m=0}^{2L} \binom{2L}{m} \frac{r^2}{r^2 +\omega^2 (L-m)^2}.
\label{eq:fidelity_stationary_value}
\end{equation}
The fidelity in the presence of resetting is plotted in Fig.~\ref{Fig:fidelity} as a function of time. We observe that the effect of a finite resetting rate consists in introducing a damping mechanism of the stable revivals characterizing the purely unitary evolution. At long times the fidelity therefore reaches a stationary value in agreement with Eq.~\eqref{eq:fidelity_stationary_value}. One observes that in the limiting case of very frequent resets the fidelity remains $1$ at all times: $\lim_{r\rightarrow +\infty}\mathcal{F}_r(t)=1$. This is the analogue of the quantum Zeno regime \cite{zeno1977one,zeno1977two} as very frequent resets do not allow the unitary evolution to alter the state from its initial condition. In practice, this is achieved whenever the condition $r\gg \omega L$ is satisfied. On the other hand, in the sparse resetting limit $r \ll \omega$, the maximum term in the sum with $m=L$ dominates over the other contributions to $\mathcal{F}_r(+\infty)$. Then one can estimate the latter as $\mathcal{F}_r(+\infty)\approx \binom{2L}{L}/2^{2L}\sim 1/\sqrt{\pi L}
$. This expressions equals the time-averaged fidelity in presence of the sole unitary evolution, consistently with what one would expect from the renewal equation for the fidelity. In general, we conclude that the convergence of the fidelity towards a time-independent value reflects the establishment of the NESS \eqref{Eq:NESS_E_basis} as a consequence of resetting. In the next sections, we show that such NESS embodies the atypical properties of the scarred states in terms of which it is constructed by showing long-range stationary correlations and sub-volume scaling of its mixedness.   

\section{Stationary state: observables and corr. functions}
\label{sec:correlations}
\begin{figure}
\centering
\includegraphics[width=1\columnwidth]{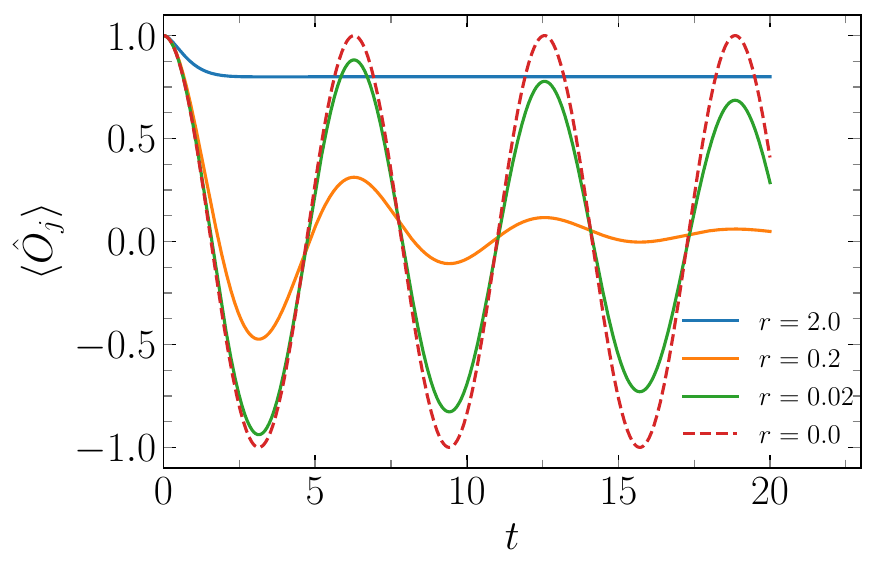}
\caption{\textbf{Dynamics of the scar quasi-particle operator as function of time}. We plot the expectation value of the observable $\hat{O}_j= \hat V_j + \hat V_j^{\dag}=e^{-i\pi j}\hat c_{j,\downarrow} \hat c_{j,\uparrow} + h.c.$ (Fermi-Hubbard) or $\hat{O}_j=e^{-i \pi j} (\hat{S}^{-}_j)^2/2 + h.c.$ (spin $1$ XY chain) as a function of time 
$t$ for the resetting dynamics interspersed with the Hamiltonian in Eqs.~\eqref{Eq:F_H_model} and \eqref{Eq:spin_1_XY_model}. The initial condition is $\ket{\alpha}$ in Eq.\eqref{Eq:coh_state} and one takes various values of the resetting rate $r=0.0$ (unitary evolution), $r=0.02$, $r=0.2$ and $r=2.0$. The other parameters are fixed as $\alpha=1.0$, $\omega=1.0$ and $L=50$.}
\label{Fig:observable}
\end{figure}

As far as the behavior of generic local observables $\hat O_j$ on a site $j$ is concerned, one has to distinguish between two cases. First one has \textit{diagonal} operators in the scar subspace, which fulfill 
\begin{equation}
\bra{\psi_m}\hat O_j \ket{\psi_n}\propto \delta_{mn}, 
\end{equation}
for any two generic scar states $\ket{\psi_{n}},\ket{\psi_{m}}$. Second, we define fully \textit{off-diagonal} operators in the scarred subspace as those operators that satisfy
\begin{equation}
\bra{\psi_m}\hat O_j \ket{\psi_n}\neq 0, \quad \mbox{only for} \quad  m\neq n.
\end{equation}
The dynamics of diagonal operators under resetting is trivial to address as their expectation value does not depend on time in presence of unitary evolution, and thus the resetting protocol cannot have any effect on their value, which equals to the following weighted average:
\begin{align}
\label{Eq:diag_obs}
    \mbox{Tr}[\hat O_j \hat\rho_r (t)] = \mbox{Tr}[\hat O_j \hat\rho(t)] &= \sum_n |c_n|^2 \bra{\psi_n} \hat O_j \ket{\psi_n},\nonumber \\
    &=\mbox{\mbox{Tr}}[\hat{O}_j \, \hat{\rho}(0)],
\end{align}
Here, the initial state $\hat{\rho}(0)$ is assumed to lie entirely within the scarred subspace, as it is the case for the coherent state $\ket{\alpha}$ in Eq.~\eqref{Eq:coh_state}. On the other hand, fully off-diagonal observables in the scarred subspace display a time dependence in absence of resetting through the dephasing induced by the mismatch between the eigenvalues of scar eigenstates $\ket{\psi_m}$, $\ket{\psi_n}$, with $m\neq n$. Therefore, stochastic resetting fundamentally alters the time evolution of the expectation values of such observables with respect to the unitary reset-free dynamics. Namely, while in absence of resetting these show persistent oscillation with a period $T\propto \omega^{-1}$, the resetting protocol forces the system to relax to the NESS \eqref{Eq:renewal_eq}, so that the aforesaid expectation values are bound to become time-independent in the $t\rightarrow +\infty$ limit. As an illustration of this behavior (see Appendix, Section \ref{S:local_obsevables}, for the detailed derivations), we consider the time evolution of the operator $\hat V_j = e^{-ikj} \hat q_j$ for the models in Eqs.~\eqref{Eq:F_H_model}-\eqref{Eq:spin_1_XY_model} and the initial condition $\ket{\alpha}$ defined in Eq.~\eqref{Eq:coh_state}. The operator $\hat q_j \in\left\{ \hat c_{j,\downarrow} \hat c_{j,\uparrow}, \frac{\left( \hat S^-_j \right)^2}{2} \right\}$ is related to the adjoint of the quasi-particle creation operator $\hat{Q}^{\dagger}_j$ in Eq.~\eqref{eq:ladder_quasiparticle} with momentum $k=\pi$. The expectation of value of this operator senses the presence of doublon/bimagnon quasi-particle excitations in the stationary states and it therefore probes the contribution from scars to the latter. The final result reads:
\begin{align}
\label{Eq:observable_vs_t}
     \mbox{Tr}[e^{-i \pi j} \hat q_j\hat\rho_r(t)] =    \left( e^{-rt+i\omega t} +r \frac{1-e^{-rt+i\omega t}}{r-i\omega}  \right)\frac{\alpha}{1+|\alpha|^2},
\end{align}
which converges to the stationary value:
\begin{align}
\label{Eq:cc}    \lim_{t\rightarrow +\infty} \mbox{Tr}[e^{-i\pi j} \hat q_j \hat\rho_r(t)] =\frac{r}{r-i\omega }  \frac{\alpha}{1+|\alpha|^2}, 
\end{align}
in the large-time limit. The result in Eq.~\eqref{Eq:cc} is easy to interpret. On the one hand, when one takes the limit $r\rightarrow +\infty$, the stationary value reads $\alpha/(1+|\alpha|^2)$, which is simply the expectation value of $e^{-i\pi j} \hat q_j$ over the initial state $\ket{\alpha}$. On the other hand, in the $r\rightarrow 0^+$ limit, the stationary value tends to zero, because since it approaches its time-averaged value in the presence of unitary time evolution only, as one can infer from Eq.~\eqref{Eq:NESS}. This value vanishes for long times.
We confirm the aforesaid observation in Fig.~\ref{Fig:observable}, where we plot the analytical result of Eq.~\eqref{Eq:observable_vs_t} plus its hermitian conjugate (in order to have an hermitian observable) for different resetting rates. We observe once again a resetting-induced damping effect on the paradigmatic infinitely long-lived oscillations observed for unitary evolution of states lying in the scarred subspace; moreover, while the damping strength increases with $r$, the stationary value approaches zero as $r$ decreases.

Going beyond mean values, we then address the effect of resetting on correlation functions. One defining feature of scarred eigenstates, see, e.g., Refs.~\cite{Moudgalya_2022,Chandran_2023,Choi_2019,Iadecola_2019,Schecter_2019,Mark_2020_2,Iadecola_2020,Gotta_2023,Sanada_2023,Imai_2025}, is the presence of off-diagonal long-range order signaled by the nonvanishing connected correlator of the quasi-particle creation operator:
\begin{align}
\lim_{L\to \infty} &\frac{1}{L^2} \braket{\psi_n| \hat{Q} \hat{Q}^{\dagger}|\psi_n}_c = \nonumber \\
&= \frac{1}{L^2} (\braket{\psi_n| \hat{Q} \hat{Q}^{\dagger}|\psi_n}-\braket{\psi_n| \hat{Q}|\psi_n} \braket{\psi_n|\hat{Q}^{\dagger}|\psi_n}) \neq 0.
\label{eq:scars_long_range}
\end{align}
This correlator is otherwise zero for thermal eigenstates in the middle of the energy spectrum. In this equation the subscript denotes the connected correlator $\braket{...}_c$ (the disconnected term $\braket{\psi_n|\hat{Q}|\psi_n}$ is anyhow zero since $\hat{Q}$ is purely off-diagonal). Equation \eqref{eq:scars_long_range} identifies scar states as condensates of quasiparticles with momentum $\pi$ (bimagnons/superconductive pairs for the spin-1 $XY$/Fermi-Hubbard model) \cite{Iadecola_2019}. Namely, Equation \eqref{eq:scars_long_range} implies that the density of quasiparticles in the momentum $\pi$ remains finite in the thermodynamic limit. In order to assess the impact of resetting on off-diagonal long-range order in the scarred subspace, we study the connected correlation function of the raising operator $\hat Q^{\dag}$ in the stationary state $\hat{\rho}_r(+\infty)$ induced by resetting:
\begin{equation} 
\label{Eq:conn_corr_fun}
    \langle \hat Q \hat Q^{\dag}\rangle_c:= \mbox{Tr}\left[\hat Q \hat Q^{\dag} \hat\rho_r(+\infty) \right]-\Biggl|\mbox{Tr}\left[\hat Q  \hat\rho_r(+\infty) \right]\Biggr|^2.
\end{equation}
\noindent Taking as a reset state an individual scar $\ket{\psi_n}$ would give a trivial evolution since this is an eigenstate. In this case, the stationary correlator in Eq.~\eqref{Eq:conn_corr_fun} would then trivially reproduce the long-range correlations \eqref{eq:scars_long_range} of the unitary dynamics. To neatly identify the effect of resetting on correlation functions, we therefore move on considering the reset state $\ket{\alpha}$ in Eq.~\eqref{Eq:coh_state}. This state, indeed, does not show off-diagonal long-range order in the thermodynamic limit for purely unitary time evolution 
\begin{equation}
 \lim_{L\to +\infty} \frac{1}{L^2}\Braket{\alpha| \hat{Q} \hat{Q}^{\dagger}|\alpha}_c =0   ,
\end{equation}
 since the state $\ket{\alpha}$ is a product state in real space, and thus it displays short-range correlations. The first term on the right hand side of Eq.~\eqref{Eq:conn_corr_fun} is the expectation value of a diagonal operator in the scarred subspace, and it is therefore time-independent as a result of the discussion in Eq.~\eqref{Eq:diag_obs}. The disconnected term on the right hand side of Eq.~\eqref{Eq:conn_corr_fun} is, instead, off-diagonal and therefore feels the effect of resetting. Namely it gets suppressed in magnitude by resetting compared to the reset-free case. Quantitatively, one finds (see Appendix \ref{S:local_obsevables} for details of the calculations) that, in the stationary limit
\begin{equation}
\mbox{Tr}\left[\hat Q \hat Q^{\dag} \hat\rho_r(+\infty)\right]  = L(L-1)\frac{|\alpha|^2}{(1+|\alpha|^2)^2}+L\frac{1}{1+|\alpha|^2}, 
\end{equation}
while 
\begin{equation}
\label{eq:ODLR_scar_reset}
\Biggl|\mbox{Tr}\left[ \hat Q^{\dag} \hat\rho_r(+\infty)\right]\Biggr|^2 = L^2 \frac{|\alpha|^2}{(1+|\alpha|^2)^2} \frac{r^2}{r^2+\omega^2},
\end{equation}
so that the final result for the connected correlation function of the quasiparticle creation operator in the stationary state takes the following stationary value:
\begin{equation}
 \lim_{L\rightarrow +\infty} \frac{1}{L^2} \langle \hat Q \hat Q^{\dag}\rangle_c =\frac{\omega^2
    }{r^2 +\omega^2}\frac{|\alpha|^2}{(1+|\alpha|^2)^2}.
\label{eq:long_range_correlations}
\end{equation}
In the infinite-resetting limit $r \to \infty$, the stationary correlator is zero. This result reflects the fact that for unitary dynamics the connected correlation function of the quasi-particle creation operator is zero in the thermodynamic limit. Any finite resetting rate $r$ leads therefore to the generation of off-diagonal long-range order and condensation of quasiparticles with momentum $\pi$ in the thermodynamic limit in the stationary state. 

As a next step we eventually compute dynamical correlation functions, where the operator $\hat{O}$ of interest is evaluated at two different times, say $t_1$ and $t_2$. 
We define the dynamical two-point correlation function of an operator $\hat{O}$ at times $t_1$ and $t_2$ ($t_2\geq t_1 \geq 0$) as:
\begin{eqnarray}
\label{Eq:dynamic_def}
C_{\hat O}(t_1,t_2)&=&\bra{\psi(0)}\hat O^{\dag}(t_2) \hat O(t_1)\ket{\psi(0)}\nonumber\\
&-&\bra{\psi(0)}\hat O^{\dag}(t_2)\ket{\psi(0)}\bra{\psi(0)}\hat O(t_1)\ket{\psi(0)},\nonumber\\     \end{eqnarray}
for an initial state $\ket{\psi(0)}$. Time evolution of operators has to be understood in the Heisenberg picture according to the Hamiltonian $\hat{H}$ in Eqs.~\eqref{Eq:F_H_model} and \eqref{Eq:spin_1_XY_model}. Since dynamical correlations involve two different times, their value in the presence of resetting cannot be obtained on the basis of the reset-free dynamics of the density matrix $\hat{\rho}(\tau)$ from the renewal equation \eqref{Eq:renewal_eq}. We, instead, directly write a renewal equation linking the dynamical correlation function $C_{r,\hat O}(t_1,t_2)$ in the presence of resetting to that one in Eq.~\eqref{Eq:dynamic_def} for unitary dynamics. In doing so, we account for the fact that observables at times $t_1$ and $t_2$ are uncorrelated whenever times $t_1$ and $t_2$ are separated by a reset event. The latter, indeed, erases all memory of the dynamics with respect to what happened before the resetting. This consideration leads to the following renewal equation for $C_{r,\hat O}(t_1,t_2)$ \cite{SO2018}:
\begin{align}
\label{Eq:dynamic}
C_{r,\hat O}(t_1,t_2) 
&= e^{-r(t_2-t_1)} \bigg[ 
   r \int_0^{t_1} d\tau \, e^{-r\tau} 
     C_{\hat O}(\tau, t_2 - t_1 + \tau) \nonumber \\
&\quad\quad + e^{-rt_1} C_{\hat O}(t_1,t_2) 
\bigg].
\end{align}
The first term accounts for the contribution of reset realizations for which the last resetting event has occurred at time $t_1-\tau$ and no further resetting event has taken place for $t_1-\tau<t<t_2$, while the second term is the contribution of trajectories for which no resetting events for $t<t_2$ have occurred. Additionally, it is worth remarking that the dynamical correlator appearing inside the integral in the first term on the right hand side of Eq.~\eqref{Eq:dynamic} is evaluated over the reset state, while the correlator $C_{\hat{O}}(t_1,t_2)$ on the second line is computed over the initial state. We take henceforth the initial and the reset state to be coincident and equal to $\ket{\alpha}$, as done throughout the manuscript and written in Eq.~\eqref{Eq:coh_state}.

In general, from Eq.\eqref{Eq:dynamic} we observe that, under the assumption that $C_{\hat O}(t_1,t_2)=C_{\hat O}(t_2-t_1)$ is a function of the time difference $t_2-t_1$ only, the function $C_{r,\hat O}(t_1,t_2)$ takes the form:
\begin{align}
\label{Eq:C_r_tt}
    &C_{r,\hat O} (t_1,t_2) = \nonumber \\
    &=e^{-r(t_2-t_1)}C_{\hat O}(t_2-t_1)\left[ r\int_0^{t_1} d\tau e^{-r\tau}  +e^{-rt_1} \right]\nonumber \\
    &=e^{-r(t_2-t_1)}C_{\hat O}(t_2-t_1).
\end{align}
This result is, for instance, satisfied under the requirement that $\ket{\psi(0)}$ in Eq.~\eqref{Eq:dynamic_def} is an eigenstate of the Hamiltonian ruling time evolution (this is a sufficient but not necessary condition). In this case, the dynamical two-point correlation function is simply weighted by the probability that no resetting occurs in the time interval $[t_1,t_2]$; the latter probability is obtained from the probability weights in the renewal equation~\eqref{Eq:dynamic}, that properly account for all stochastic realizations featuring no stochastic reset for time $t\in [t_1,t_2]$. 
This equation shows that the system becomes completely uncorrelated for time separations $(t_2-t_1) \gg 1/r$ due to the fact that a reset occurs with high probability in such a time interval. Differently from the case of spatial correlations of Eq.~\eqref{eq:long_range_correlations}, we therefore conclude that stochastic resetting damps the long-range time crystalline ordering typical of scarred eigenstates. As an example of such result, we compute the dynamical correlation function of the quasi-particle creation operator $\hat Q^{\dag}$ generating the tower of quantum many-body scars. In this case, we obtain (see Appendix~\ref{S:dyn_corr_fun} for the details of the analysis)
\begin{align}
  C_{r,\hat Q^{\dag}} (t_1,t_2) &= e^{-r(t_2-t_1)}   
  C_{\hat Q^{\dag}} (t_1,t_2) =\\
  &=e^{-r(t_2 - t_1) +i\omega(t_2-t_1)} L \frac{1}{(1+|\alpha|^2)^2}.  
\label{eq:correlation_function_Q_dyn_final}
\end{align}
We observe that this result for $t_1=t_2=t$ reduces to the average over the reset realizations (denoted with the overline $\overline{x}$ symbol) of the connected correlation function 
\begin{equation}
\overline{\Braket{\psi(0)|\hat{Q}^2(t)|\psi(0)}-\Braket{\psi(0)|\hat{Q}(t)|\psi(0)}^2},
\end{equation}
at time $t$. This result is different from that one obtained by subtracting the average over resetting of the connected $\overline{\Braket{\psi(0)|\hat{Q}^2(t)|\psi(0)}}$ and disconnected $\overline{\Braket{\psi(0)|\hat{Q}(t)|\psi(0)}}^2$  part of the correlation function, where the average over realizations is taken on each of the two terms separately. This follows from the nonlinearity of the disconnected term. This is the reason why Eq.~\eqref{Eq:conn_corr_fun} differs from Eq.~\eqref{eq:correlation_function_Q_dyn_final}.

\section{Mixedness logarithmic stationary law}
\label{sec:entropy_mix}

In this section, we characterize the NESS obtained via resetting by quantifying the associated mixedness. 
As mentioned, for the sake of treating a problem that is potentially within reach of experimental realization, we consider the family of initial states introduced in Eq.\eqref{Eq:coh_state}. These are coherent states in the scarred subspace, that feature the property of being product states in real space. For a generic density matrix $\hat\rho$, we measure its mixedness by evaluating its $n$-Rényi entropy:
\begin{equation}
\label{Eq:n_renyi}
    S_n(\hat\rho)= \frac{1}{1-n}\log\left(\mbox{Tr}[\hat \rho^n] \right),
\end{equation}
as well as the von Neumann entropy:
\begin{equation}
\label{eq:vNeumann_entropy}
    S_{vN}(\hat\rho)= -\mbox{Tr}[\hat\rho\log \hat\rho]= \lim_{n\rightarrow 1} S_n(\hat\rho).
\end{equation}
The mixedness of the density matrix $\mbox{Tr}[\hat{\rho}^2]$ is, namely, related to the second Rényi entropy, which is equal to zero for a pure state ($\mbox{Tr}[\hat{\rho}^2]=1$). Since unitary dynamics preserves purity of quantum states, any non zero value for the Rényi/von Neumann entropies in Eqs.~\eqref{eq:vNeumann_entropy} and \eqref{Eq:n_renyi} is generated by the stochastic resetting protocol, which leads the dynamics for any finite resetting rate $r$ towards the mixed stationary state in Eq.\eqref{Eq:NESS}. It is also important to remark that resetting dynamics with Poissonian waiting time distribution \eqref{eq:Poissonian} is equivalent to dissipative dynamics of the Lindblad form \cite{Hartmann2006,Linden2010,Rose2018spectral,Perfetto_2021}. In the latter case, Rényi/von Neumann entropies do not quantify entanglement. They, instead, quantify mixedness generated by the resetting protocol as explained above. 

The reset-free evolution from the state $\ket{\alpha}$ is obtained as  
\begin{equation}
\hat\rho(\tau)=\ket{\alpha(\tau)}\bra{\alpha(\tau)}, \quad \mbox{and} \quad \ket{\alpha(\tau)}=\ket{\alpha e^{i\omega \tau}}.
\label{eq:coherent_state_evolution}
\end{equation}
One therefore finds that the states $\ket{\alpha(\tau)}$ are product states in real space at all times $\tau$. This allows to obtain an exact expression for the $n$-Rényi entropy of their corresponding time-evolved state. The stationary density matrix has the form 
\begin{equation}
\hat\rho_r(+\infty) = r\int_0^{+\infty} d\tau e^{-r\tau} \ket{\alpha(\tau)}\bra{\alpha(\tau)}.
\label{eq:ness_alpha}
\end{equation}
We note that this state is a convex combination of product states and as such does not possess any entanglement, as quantified by bipartite concurrence \cite{concurrence}. It, however, possesses mixdness that we quantify by evaluating the $n$-Rényi entropy of the stationary density matrix $\hat\rho_r(+\infty)$. Taking the trace of both sides of Eq.~\eqref{eq:ness_alpha} leads to the general expression (the derivation is reported in Appendix \ref{App:steady_S2}):
\begin{align} 
\label{Eq:n_Rényi_coh_states}
&\mbox{Tr}[\hat\rho^n_{r}(+\infty)] = e^{(1-n)S_n(\hat \rho_{r}(+\infty))}=  \nonumber \\
& \!|\mathcal{N}_{L}|^{2n} r^n \!\!\!\!\!\! \sum_{l_1,\dots, l_n=0}^{L}\!\left(\prod_{j=1}^n \binom{L}{l_j} |\alpha|^{2 l_j}\! \right)\!\!\left(\prod_{k=1}^n \!\frac{1}{r-i\omega(l_{k-1}-l_k)} \!\right)\!,
\end{align}
where $|\mathcal{N}_L|^2=1/(1+|\alpha|^2)^L$ is the normalization of the coherent state in Eq.~\eqref{Eq:coh_state} and, in the last product of the above expression, we imply the identification $l_0=l_n$. Let us now specialize for simplicity to the case $n=2$, which directly link to mixedness of the stationary state:
\begin{align}
\label{Eq:2_Renyi}
    &e^{-S_2(\hat\rho_{r}(+\infty))} = \nonumber \\
    &\frac{1}{(1+|\alpha|^2)^{2L}}\sum_{m,n=0}^{L} \binom{L}{m}\binom{L}{n} |\alpha|^{2(m+n)}\frac{r^2}{r^2+\omega^2(m-n)^2}.
\end{align}
\begin{figure}
\centering
\includegraphics[width=1\columnwidth]{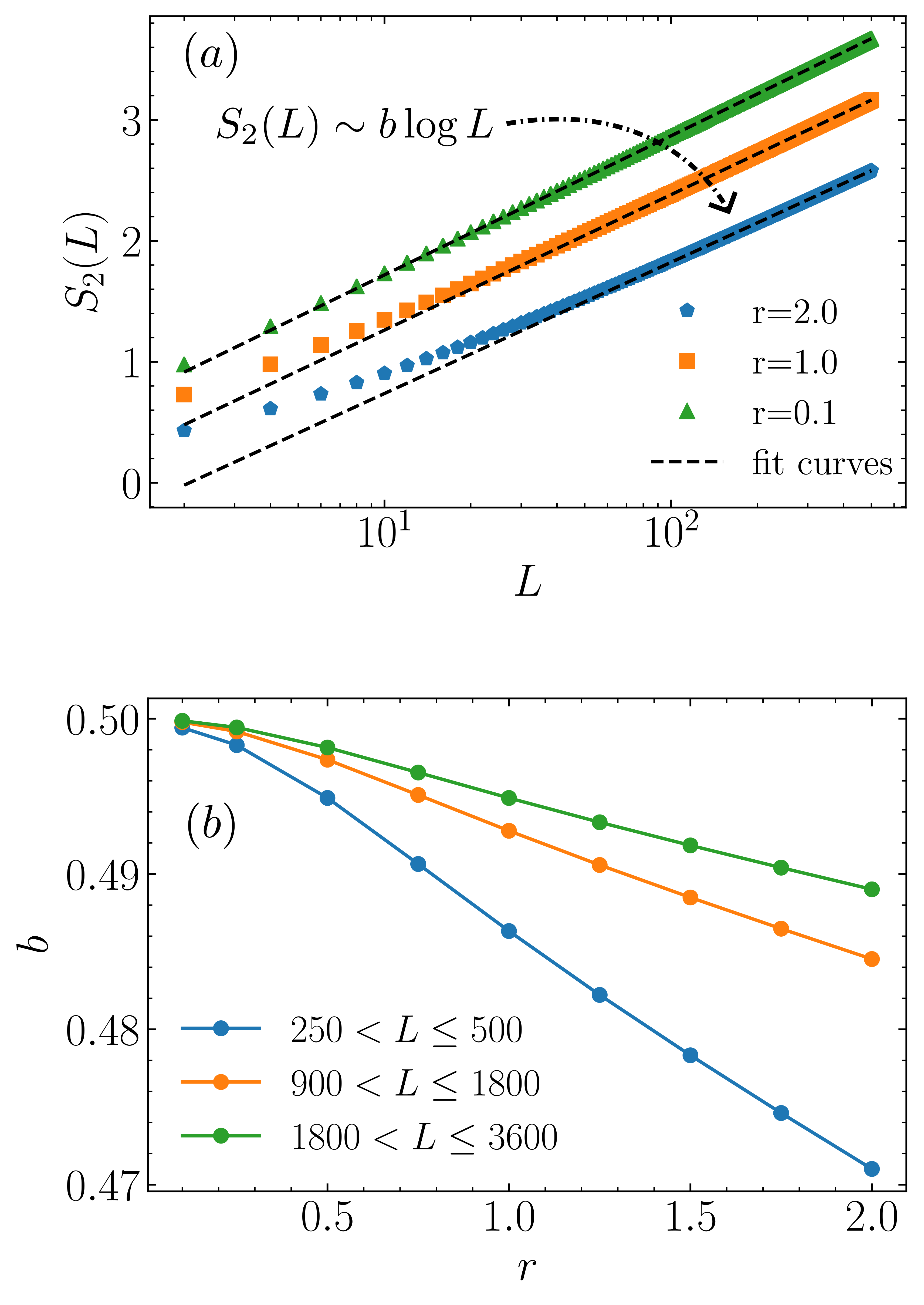}
\caption{(a) \textbf{Logarithmic scaling of Rényi-$2$ entropy as a function of size $L$}. The parameters are $\omega = 2.0$ and $\alpha = 1.0$, while various values of the resetting rate $r$ are reported. Data are obtained from the numerical evaluation of the exact formula of Eq.~\eqref{Eq:2_Renyi}. The dashed curves are the results of fitting to a function form $f(x)=a+b\log(x)$, where $a$ and $b$ are fitting parameters. The fitting has been performed for $250<L\leq 500$. (b) Fitting parameters $b$ as a function of the resetting rate $r$ for different ranges of system size $L$ where the fitting is performed. We see that as the resetting grows, one needs to consider larger sizes for the fitting region in order for the fitting parameter $b$ to converge to the theoretical value $1/2$ of Eq.~\eqref{eq:entropy_asymptotic_scaling}.}
\label{Fig:entropy_scaling}
\end{figure}

\begin{figure*}
\centering
\includegraphics[width=\textwidth]{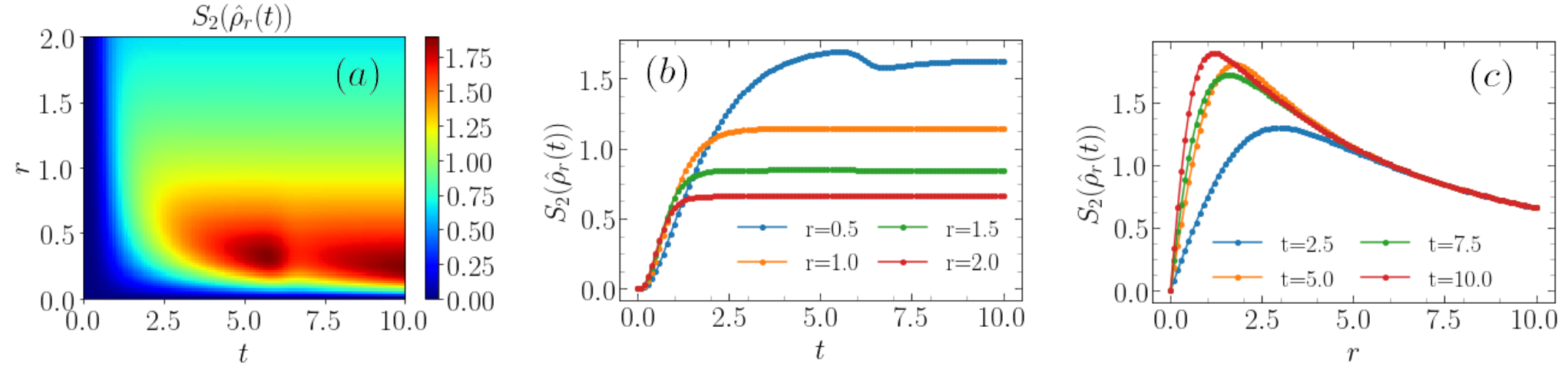}
\caption{(a): \textbf{Time evolution of the Rényi-$2$ entropy.} The parameters are $\omega = 1.0$, $\alpha = 2.0$
and $L=30$ for all the panels. (a) Density plot of the Rényi-$2$ entropy as a function of time $t$ and resetting rate $r$. (b): Rényi-$2$ entropy as a function of time $t$ for various values of the resetting rate $r$. Mixedness of the stationary state is maximized for small resetting rate. (c): Rényi-$2$ entropy as a function of $r$ for various values of the time $t$. The maximum of $S_2(\hat\rho_{r}(t))$ is generated by the fact that both for $r\to 0^{+}$ (approximately unitary dynamics) and $r\to \infty$ (quantum Zeno regime) the state is pure. }
\label{Fig:4}
\end{figure*}
In the strong resetting limit $r\to \infty$, the r.h.s. of Eq.~\eqref{Eq:2_Renyi} tends to one, and thus one gets that $\lim_{r\rightarrow +\infty} S_2(\hat\rho_{r}(+\infty))=0$, as expected, since a high resetting rate does not allow for any time evolution to occur and forces the state into its initial condition $\ket{\alpha}$, which is a pure state. In order for the strong resetting limit to set in at finite size $L$, one must be able to perform the approximation $r^2/[r^2 + \omega^2(m-n)^2]\approx 1$, i.e., $r\gg \omega|m-n|$, for all values of $m,\, n$. Therefore, the strictest condition to be satisfied in order to reach the strong resetting limit at size $L$ is $r\gg \omega L$. On the other hand, in the sparse resetting limit $r\to 0^{+}$ of Eq.~\eqref{Eq:2_Renyi}, one has that, under the condition $r\ll \omega$, the approximate value of $S_2(\hat\rho_{r}(+\infty))$ is obtained by retaining only the terms with $m=n$ in the summation in Eq.~\eqref{Eq:2_Renyi}, as the latter are of zeroth order in the resetting rate $r$, while all terms with $m\neq n$ are suppressed by a factor $r$. The physical meaning of this approximation is that the small $r$ limits suppresses off-diagonal elements ($m\neq n$) in the stationary density matrix written in the scar energy subspace driving the system towards the diagonal ensemble of Eq.~\eqref{eq:diagonal_ensemble}. The resulting expression for the $2$-Rényi entropy of the half-chain reduced density matrix is obtained in Appendix \ref{app:long_L_2renyie} and it is equal to
\begin{align}
\label{Eq:2_renyi_small_r}
    \lim_{r\rightarrow 0^+} S_2(\hat\rho_{r}(+\infty)) = -\log\left[\frac{1}{(1+|\alpha|^2)^{2L}}\sum_{n=0}^{L}\binom{L}{n}^2 |\alpha|^{4n}\right].
\end{align}
In the large-$L$ limit, Eq.~\eqref{Eq:2_renyi_small_r} can be further estimated via saddle-point methods to obtain the following scaling:
\begin{align}
   \lim_{r\rightarrow 0^+} S_2(\hat\rho_{r}(+\infty)) &\approx    \log\left[ \frac{2|\alpha| \sqrt{\pi L}}{1+|\alpha|^2}\right] \nonumber \\
   &\sim \frac{1}{2}\log L \quad \mbox{as} \quad L\to \infty,
\label{eq:entropy_asymptotic_scaling}
\end{align}
In Fig.~\ref{Fig:entropy_scaling}, we compare the exact Rényi second entropy computed from Eq.~\eqref{Eq:2_Renyi} with a logarithmic asymptotics $S_2(L)=a+b\log(L)$, with $a,b$ fitting parameters. In Fig.~\ref{Fig:entropy_scaling}(a), we observe that such logarithmic scaling excellently reproduces the large size limit not only for small values of the resetting rate ($r=0.1$), but also for finite ones ($r=1$ and $2$). We show analytically in Appendix \ref{app:long_L_2renyie}, that the result \eqref{eq:entropy_asymptotic_scaling} applies, indeed, for generic and finite values of the resetting rate $r$. In Fig.~\ref{Fig:entropy_scaling}(b),  we plot the fitting parameter $b$ as a function of the resetting rate for different values of the system size $L$. We find that $b$ approaches the value $1/2$, in agreement with our asymptotic result \eqref{eq:entropy_asymptotic_scaling}. The value of $L$ where such asymptotic scaling applies grows as $r$ gets bigger. In Appendix \ref{app:long_L_n_renyie}, we further show that the logarithmic scaling \eqref{eq:entropy_asymptotic_scaling} applies generically to all the Rényi entropies with index $n\geq 3$.

The result in Eq.~\eqref{eq:entropy_asymptotic_scaling} reproduces the half-chain entanglement entropy of the scarred states $\ket{\psi_n}$. The latter is, indeed, known to display entanglement entropy whose scaling in the system size is logarithmic \cite{EE_scars_FH,Iadecola_2019,Schecter_2019,Iadecola_2020,Shibata_2020}. This entanglement content is atypically low compared to that of thermal eigenstates in the middle of the energy spectrum whose entanglement entropy is extensive in the system size $L$ (volume law). We can understand the logarithmic scaling of the stationary state mixedness induced by resetting (the details of the calculations are reported in Appendix \ref{app:rsmall_reduced_rho}) by computing the half-chain density matrix $\hat{\rho}_{L/2}(\ket{\psi_n}\bra{\psi_n})$ associated to a scar eigenstate $\ket{\psi_n}$ (we assume here $n\leq L/2$ for the sake of illustration purposes): 
\begin{equation}
\hat{\rho}_{L/2}(\ket{\psi_n}\bra{\psi_n})=\sum_{m=0}^{n}p_n\ket{\psi_{m,L/2}}\bra{\psi_{m,L/2}}.
\label{eq:reduced_state_scar}
\end{equation}
Here $\ket{\psi_{m,L/2}}$ is a scarred eigenstate for the half-chain system defined in $j=1,2\dots L/2$. The state \eqref{eq:reduced_state_scar} has the same structure (of course on the smaller Hilbert space associated to a half chain) as the diagonal ensemble \eqref{eq:diagonal_ensemble} obtained in the weak resetting limit. Namely, the coefficients $p_n$ show the same scaling as a function of the system size $L$ as the coefficients $c_n$ in Eq.~\eqref{eq:diagonal_ensemble}. This eventually causes the mixedness of the stationary state in the presence of resetting to display the same scaling as the entanglement entropy of a scar eigenstate. This result has interesting physical implications since it shows that the application of stochastic resetting, a classical stochastic process, onto a quantum scarred dynamics can be exploited to prepare the reduced density matrix \eqref{eq:reduced_state_scar} that corresponds to highly-entangled quantum many-body states. 

We eventually conclude this section by studying the dynamics in time of the $2$-Rényi entropy. The calculation is carried out with a very similar technique to the one presented for obtaining the stationary-state $2$-Rényi entropy, leading to an exact expression, which is shown in the Appendix \ref{app:time_dependence_Renyie}. The results are reported in Fig.~\ref{Fig:4}. The density plot in Fig.~\ref{Fig:4}(a) shows that the time needed to reach stationarity increases as the resetting rate is reduced $r\rightarrow 0^+$. This information is equivalently displayed in Fig.~\ref{Fig:4}(b), where interestingly we can observe that the stationary mixedness increases as the resetting rate decreases. This can be interpreted by noting that the stationary state for low resetting corresponds to the reduced density matrix of an entangled scar state according to Eq.~\eqref{eq:reduced_state_scar}. The latter has a significant entanglement content, which reflects into the mixedness of the obtained stationary state. For finite $r$, instead, the reduced state obtained from the resetting NESS does not map to a scar eigenstate and its entanglement is consequently reduced due to the classical noise from resetting. This eventually translates into a lower stationary mixedness for the resetting stationary state.   
As a result, when plotting $S_2(\hat\rho_{r}(t))$ as a function of $r$ for several values of $t$ in Fig.~\ref{Fig:4}(c), one observes that the Rényi-$2$ entropy becomes stationary for all values of $r$ such that $r^{-1}<t$. The position of the maximum as a function of $r$ shifts to lower and lower values as the time $t$ grows, which again reflects that the maximal mixedness for the stationary state is attained in the weak resetting limit.

\section{Weak resetting and concentration around single eigenstate}
\label{sec:6equivalence}
In the previous section we proved that the stationary state \eqref{eq:diagonal_ensemble} in the weak resetting limit has the same structure as the reduced density matrix \eqref{eq:reduced_state_scar} associated to a scarred eigenstate. We now show that both these mixed states in the thermodynamic limit $L\to \infty$ show concentration around a single scarred eigenstate $\ket{\psi_{n^*}}$. The quantum number $n^{*}$ of the selected eigenstate is determined only by the parameter $\alpha$ labelling the coherent state and thus can be controlled with the reset protocol. 

Concentration around a single scar eigenstate simply follows by writing the expression for the weights $c_n$ identifying the diagonal ensemble  (Appendix \ref{S:local_equivalence}, for the detailed derivation)
\begin{equation}
|c_n|^2 =\binom{L}{n} (y^{\ast}(\alpha))^n (1-y^{\ast}(\alpha))^{L-n}.
\label{eq:weigths_binomial_main}
\end{equation}
This formula identifies a binomial distribution with mean value $y^{\ast}(\alpha)$ given by 
\begin{equation}
\label{Eq:n_alpha_mean}
    y^*(\alpha) = \frac{|\alpha|^2}{1+|\alpha|^2}.
\end{equation}
In the thermodynamic limit, the binomial distribution concentrates around the mean value \eqref{Eq:n_alpha_mean} according to the law of large numbers. This implies that the diagonal ensemble \eqref{eq:diagonal_ensemble} concentrates around a single scarred state with quantum number $n^{\ast}= L y^{\ast}(\alpha)$ and therefore for every local observable $\hat{O}$ we have the equivalence:
\begin{equation}\label{Eq:equivalence}
    \lim_{L\rightarrow +\infty} \lim_{r\rightarrow 0^+} \mbox{Tr}[\hat O \hat \rho_r(+\infty;\alpha)] = \bra{\psi_{\lfloor Ly^*(\alpha)\rfloor}}\hat O\ket{\psi_{\lfloor Ly^*(\alpha) \rfloor}}.
\end{equation}
It is important to emphasize that this equation applies to local observables only, i.e., to operators $\hat{O}$ such that their expectation value $\braket{\psi_n|\hat{O}|\psi_n}$ over the scarred eigenstates is not exponentially large in the system size $L$. As a matter of fact Eq.~\eqref{Eq:equivalence} is obtained by saddle point analysis, which selects the optimal scarred state $n^{\ast}$. In the case of an operator $\hat{O}$ whose scar expectation value is exponential in the system size, the saddle point determining the thermodynamic limit would get a shift compared to $n^{\ast}$ and therefore Eq.~\eqref{Eq:equivalence} does not hold (see Appendix \ref{S:local_equivalence}). We give an example of this aspect at the end of this section. We note that the local equivalence expressed by Eq.~\eqref{Eq:equivalence} is in agreement with the result of Ref.~\cite{Morettini_2025} (cf. Eq.~(A8) therein), where expectation values of local observables over scarred eigenstates are shown in the thermodynamic limit to be equal to the time average over a period $T=2\pi/\omega$ of the expectation value over the coherent state $\ket{\alpha(t)}$ of Eq.~\eqref{eq:coherent_state_evolution}. This statement is clearly equivalent to our result since time average of the quench dynamics from the coherent state $\ket{\alpha(t)}$ generates nothing but the diagonal ensemble. Stochastic resetting thus offers a neat protocol to dynamically generate this equivalence and to exploit it to prepare local properties of scarred eigenstates. 

\begin{figure}[t]
\centering
\includegraphics[width=1\columnwidth]{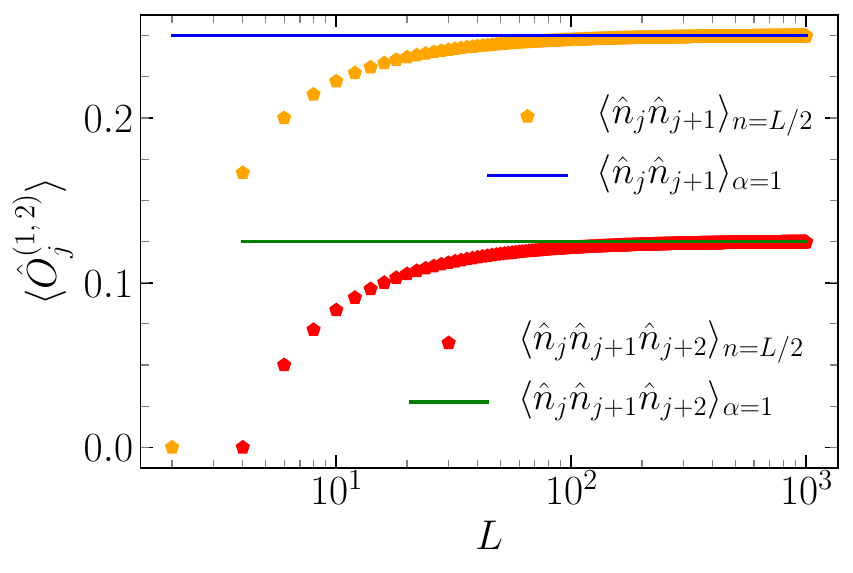}
\caption{\textbf{Local equivalence between the diagonal ensemble and a single scar for weak resetting.} Parameters are $\alpha=1.0$ and arbitrary $\omega\in \mathbb{R}$. Solid lines represent the expectation value of $\hat O_j^{(1)} = \hat n_j \hat n_{j+1}$ and $\hat{O}_j^{(2)}=\hat n_j \hat n_{j+1} \hat{n}_{j+2}$ over the stationary mixed state $\hat\rho_r (+\infty)$ with $r=0^+$, in Eqs.~\eqref{eq:O_1_left} and \eqref{eq:O_l_2}, respectively. Dotted lines represent the expectation values of the same operators over the scar eigenstate with quasiparticle number $n^{\ast}=L/2$, Eqs.~\eqref{eq:O1_right} and \eqref{eq:O2_r}, respectively.}
\label{Fig:local_equiv}
\end{figure}

We benchmark this result by computing the two sides of Eq.~\eqref{Eq:equivalence} for $\alpha=1$ and for the diagonal observables $\hat{O}_j^{(1)}=\hat n_j \hat n_{j+1}$ and $\hat{O}_j^{(2)}=\hat n_j \hat n_{j+1} \hat n_{j+2}$, realized via the tensor product of the projectors $\hat n_l$ onto the quasiparticle states on sites $l$. The operators $\hat{O}_j^{(1)}$ and $\hat{O}_j^{(2)}$ thus measure the probability of finding two (on sites $j$ and $j+1$) and three neighboring (on sites $j,j+1$ and $j+2$) excitations, respectively. In the case of the tower of $\eta$-pairing states in the Fermi-Hubbard model \eqref{Eq:F_H_model}, one has
$\hat n_j = \hat n_{j,\uparrow} \hat n_{j,\downarrow}$, while for the $XY$ chain \eqref{Eq:spin_1_XY_model} and the spin-$1$ tower of bimagnonic states, one may write $\hat n_j = \ket{+1}_j \bra{+1}_j$. For the expectation value of the operator $\hat{O}_j^{(1)}$ over $\hat\rho_r(+\infty)$ the diagonal ensemble in the weak resetting limit one has:
\begin{equation}
   \lim_{r\rightarrow 0^+} \mbox{Tr}\left[\hat n_j \hat n_{j+1} \hat \rho_{r}(+\infty)\right] = \frac{1}{2^L}\sum_{n=2}^L \binom{L}{n} \frac{\binom{L-2}{n-2}}{\binom{L}{n}} = \frac{1}{4}.
\label{eq:O_1_left}
\end{equation}
The right hand side of Eq.~\eqref{Eq:equivalence} can also be readily computed for the scarred state $n^{\ast}=L/2$ ($\alpha=1$) and it gives 
\begin{equation}
\bra{\psi_{\lfloor L/2\rfloor}}\hat n_j \hat n_{j+1}\ket{\psi_{\lfloor L/2 \rfloor}} = \frac{\lfloor L/2 \rfloor (\lfloor L/2 \rfloor -1) }{L(L-1)}. 
\label{eq:O1_right}
\end{equation}
The expressions \eqref{eq:O_1_left} and \eqref{eq:O1_right} coincide in the large-$L$ limit, as predicted. Similarly, in the case of the observable $\hat{O}_j^{(2)}=\hat n_j \hat n_{j+1} \hat n_{j+2}$, one has the expectation value over the diagonal state
\begin{equation}
    \lim_{r\rightarrow 0^+}\mbox{Tr}\left[\hat n_j \hat n_{j+1}\hat n_{j+2} \hat \rho_{r}(+\infty)\right]  = \frac{1}{2^L}\sum_{n=3}^L \binom{L}{n} \frac{\binom{L-3}{n-3}}{\binom{L}{n}} = \frac{1}{8}.
    \label{eq:O_l_2}
\end{equation}
The expectation value over the scar eigenstate $n^{\ast}=L/2$ reads as 
\begin{align}
\bra{\psi_{\lfloor L/2\rfloor}}\hat n_j \hat n_{j+1}&\hat n_{j+2}\ket{\psi_{\lfloor L/2 \rfloor}}= \nonumber \\
&= \frac{\lfloor L/2 \rfloor (\lfloor L/2 \rfloor -1) (\lfloor L/2 \rfloor -2)}{L(L-1)(L-2)},
\label{eq:O2_r}
\end{align}
and coincides with Eq.~\eqref{eq:O_l_2} in the large-$L$ limit. In Fig.\ref{Fig:local_equiv}, we plot Eqs.~\eqref{eq:O_1_left} and \eqref{eq:O1_right} as a function of $L$ for the operator $\hat{O}_j^{(1)}$ and Eqs.~\eqref{eq:O_l_2} and \eqref{eq:O2_r} for the operator $\hat{O}_j^{(2)}$. In both the cases, we clearly observe that in the thermodynamic limit $L\rightarrow +\infty$ expectation values over the scar eigenstate $n^{\ast}=L/2$ are locally indistinguishable from those computed over the diagonal ensemble describing the resetting NESS according to Eq.~\eqref{Eq:equivalence}.

As stated above, the equivalence in Eq.~\eqref{Eq:equivalence} holds only for local observales. In order to see this, we can consider the fidelity of the state $\ket{\alpha=1}$. This amounts to considering the expectation value of the nonlocal operator $\ket{\alpha}\bra{\alpha}$. In this case, the left hand side of Eq.~\eqref{Eq:equivalence} reads
    \begin{align}     
    \mbox{Tr}[\ket{\alpha}\bra{\alpha}\hat{\rho}_0(+\infty)]&=\bra{\alpha=1}\hat\rho_0(+\infty)\ket{\alpha=1}\nonumber \\
    &= \sum_n |c_n|^2 |\braket{\alpha=1 | \psi_n}|^2 \nonumber \\
    &= \sum_n |c_n|^4 \underset{L\gg 1}{\approx} \frac{1}{\sqrt{\pi L}},  
\label{eq:left_fidelity}
\end{align}
where the last step follows from a saddle-point evaluation of the sum, as reported in Appendix \ref{S:local_equivalence}. On the other hand, the fidelity of the state $\ket{\alpha}$ with the representative scar $\ket{\psi_{n^{\ast}}}$ ($n^{\ast}$=$L/2$) gives the right hand side of Eq.~\eqref{Eq:equivalence}. This leads to
\begin{align}
\braket{\psi_{L/2}|\ket{\alpha}\bra{\alpha}|\psi_{L/2}}&=\Bigl|\braket{\alpha=1 | \psi_{L/2}}\Bigr|^2 \nonumber \\
&= \frac{1}{2^L}\binom{L}{L/2}\underset{L\gg 1}{\approx}\sqrt{\frac{2}{\pi L}}.
\label{eq:right_fidelity}
\end{align}
The discrepancy by of factor $2$ between Eq.~\eqref{eq:left_fidelity} and \eqref{eq:right_fidelity} is caused by the fact that the overlap $|\braket{\alpha|\psi_n}|^2$ in Eq.~\eqref{eq:left_fidelity} depends on the exponential of the system size $L$. This leads to a shift of the saddle point, which in this specific case gives a factor two since $|c_n|^2=\braket{\alpha|\psi_n}$ and therefore the action determining the stationary fidelity is doubled.

\section{Discussion and outlook}
\label{sec:conclusions}
In this work we discussed the effect of stochastic resetting onto the unitary many-body dynamics arising from interacting Hamiltonians featuring towers of many-body quantum scars. Namely, we considered both spin-1 $XY$ Hamiltonian \eqref{Eq:spin_1_XY_model} and deformed Fermi-Hubbard models \eqref{Eq:F_H_model}. We considered the case where the initial and reset states coincide and they are contained entirely within the scarred subspace, as it is the case for the coherent-like states in Eqs.~\eqref{Eq:coh_state}. In both the cases, the resetting protocol eventually damps the long-lived oscillations arising from the unitary evolution within the scarred manifold and it drives the system towards a NESS. We characterized the latter by computing the dynamics of both the fidelity with respect to reset state and that of local observables. For correlation functions we interestingly observe that stochastic resetting induces spatial off-diagonal long-range order (cf. Eq.~\eqref{eq:long_range_correlations}) in the NESS. Time-crystalline order in dynamical correlation functions is, instead, damped by resetting since this decorrelates time intervals separated by resetting events (see Eq.~\eqref{Eq:C_r_tt}). We then computed the mixedness, as quantified by the Rényi-$2$ entropy, generated by the resetting protocol as a function of time and also in the NESS eventually attained. This quantity shows a striking connection to the entanglement content of scarred eigenstates. Specifically, we find in Fig.~\ref{Fig:entropy_scaling} that the stationary mixedness displays a logarithmic scaling as a function of the system size $L$, which directly reflects the same scaling shown by the entanglement of scarred eigenstates \cite{Iadecola_2019,Schecter_2019,Iadecola_2020,Shibata_2020}. We clarified this surprising feature by considering the weak resetting regime. In this limit, the NESS density matrix is a diagonal ensemble and it has the same structure as the reduced density matrix associated to a scar eigenstate. In addition, we also show that such mixed density matrix displays dominant weight in the thermodynamic limit around a single scarred eigenstate. This state is uniquely fixed by the chosen value of the complex parameter $\alpha$ parametrizing the reset state. The stationary state obtained in the presence of weak resetting is thus locally indistinguishable from a pure scarred eigenstate according to Eq.~\eqref{Eq:equivalence}. This is crucially achieved by interspersing the unitary dynamics with a small number of resets to product states $\ket{\alpha}$, which are simple to prepare. This analysis might thus represent a viable way to prepare local properties of entangled scarred states. The advantage of the present proposal to prepare local properties of entangled scarred states is that it does not require any postselection of the quantum trajectories induced by stochastic realizations of resetting. On the other hand, however, the weak resetting limit implies that the time needed to reach stationarity might be too long. 

In this sense, an interesting future direction would be to accelerate convergence to stationarity by applying resetting with a finite resetting rate for an initial transient time, as in the protocols studied in Refs.~\cite{bao_accelerating,solanki2025universal}. A natural extension of this work requires studying the stability of our results to dissipation and measurements, and therefore considering dissipative dynamics. It is then relevant to study the interplay between scarred dynamics and conditional reset protocols \cite{Perfetto_2021,Magoni_2022,Anish_conditional,solanki2025universal}, where multiple reset states can be chosen depending on the outcome of a measurement. This can be specifically interesting for dissipative dephasing dynamics hosting multiple stationary states \cite{Marche_2025} both of scarred and non scarred type. In these models, the application of conditional protocols is expected to lead to a mixture of such features and therefore to collective athermal behavior. It would be relevant to formulate the present dynamics at stroboscopic discrete times, as in Floquet driving \cite{haldar2022statistical,Lazarides_floquet}. This amounts to studying the dynamics dictated by unitary gates interspersed with resets \cite{wald2025stochastic} and could allow to directly test our predictions on state-of-the-art quantum hardware devices. 

\section*{Acknowledgments} L.G. thanks Luca Capizzi and Gianluca Morettini for enlightening discussions. This work is supported by the Swiss National Science Foundation under Division II
(Grant No. 200020-219400). M. K. acknowledges the support of the Department of Atomic Energy, Government of India, under project no. RTI4001. G.P. acknowledges funding from the Swiss National Science Foundation (SNSF) through Grant No.~10005336.

\bibliography{reset_bib}

\newpage
\onecolumngrid
\appendix

\section*{Appendix}
In the Appendix we provide the details behind the calculations that lead to the results presented in the main text. In all the following sections, we consider Hamiltonian hosting towers of quantum many-body scars $\ket{\psi_n}$ generated by the repeated application of a raising operator $\hat Q^{\dag}$ to a reference state $\ket{\psi_0}$ according to Eq.~\eqref{eq:scars_towers}. We further assume the form 
\begin{equation}
\hat Q^{\dag}=\sum_{j=1}^L e^{ikj} \hat q^{\dag}_j, \quad \ket{\psi_n} \propto (\hat{Q}^{\dagger})^n \ket{\psi_0},
\label{eq:quasi_particle_operator}
\end{equation}
for the raising operator, $L$ being the system size and (i) $\left(\hat q^{\dag}_j\right)^2 =0$, (ii) $[\hat q^{\dag}_j, \hat q^{\dag}_l]=0$, $\forall j, l$. From the condition in Eq.~\eqref{Eq:sga} defining the spectrum generating algebra, we also have that the eigenvalues of the scarred eigenstates are equispaced 
\begin{equation}
\hat H (\hat Q^{\dag})^n \ket{0}=-\omega n (\hat Q^{\dag})^n \ket{0}, \quad n=0,1,2\dots L,
\end{equation}
for some $\omega\in\mathbb{R}$. These hypothesis are satisfied by both the Fermi-Hubbard and spin-1 $XY$ chain in Eqs.~\eqref{Eq:F_H_model} and \eqref{Eq:spin_1_XY_model}, respectively. In the former case, the tower of many-body quantum scars is given  by $\eta$-pairing states:
\begin{equation}
\label{S:Eq:eta_pair_tower}
    \ket{\psi_n} = \frac{1}{n!\sqrt{\binom{L}{n}}} \left(\sum_j (-1)^j \hat c^{\dag}_{j,\uparrow} \hat c^{\dag}_{j,\downarrow} \right)^n \ket{0}.
\end{equation}
while in the latter by bimagnon excitations
\begin{equation}\label{S:Eq:spin_1_tower}
    \ket{\psi_n} = \frac{1}{n!\sqrt{\binom{L}{n}}} \left(\sum_j \frac{(-1)^j}{2} (\hat S^+_j)^2\right)^n \bigotimes_{j=1}^L \ket{-1}_j,
\end{equation}
as detailed in Sec.~\ref{sec:model} of the main text. For the Fermi-Hubbard model the energy spacing between consecutive scars is given by $\omega=2\mu -U$, while for the spin chain by $\omega=-2h$. All the results that we present henceforth in the Appendix apply to both class of models once the model-specific form of the spacing $\omega$ and the quasi-particle operator $\hat{Q}^{\dagger}$ is fixed. We therefore keep these quantities as general in the following. 

\section{Fidelity}
\label{S:fidelity}
We here provide details about the calculation of the fidelity in Eq.~\eqref{eq:fidelity_stationary_value} of Sec.~\ref{sec:fidelity}. From the renewal equation:
\begin{equation}
\label{S:Eq:renewal_eq}
    \hat \rho_r(t) = e^{-rt}\hat\rho(t) + r\int_0^t d\tau \, e^{-r\tau} \hat \rho(\tau),
\end{equation}
we can write the density matrix $\hat{\rho_r}(t)$ evolved in the presence of resetting in terms of the reset-free unitary evolution $\hat{\rho}(t)$. We are considering here the exponential waiting-time distribution \eqref{eq:Poissonian} and we initialize the system in a state entirely $\ket{\psi(0)}$ contained within the scarred manifold $\ket{\psi_r}$ 
\begin{equation}
\ket{\psi_r}=\ket{\psi(0)}=\sum_n c_n \ket{\psi_n}.
\label{eq:scars_initial}
\end{equation}
In this equation we further assumed, without loss of generality, that the initial state coincides with the reset state $\ket{\psi_r}=\ket{\psi(0)}$. This is not a restrictive assumption since the initial state determines only the transient dynamics in the renewal equation \eqref{S:Eq:renewal_eq}, but does not impact on the stationary state. The renewal equation after decomposition of the reset-free density matrix in terms of the scarred states reads as:
\begin{equation} \label{S:Eq:generic_rho_r}
    \hat\rho_r(t) = \sum_{m,n} c_m c_n^* \left\{e^{-[r+i(E_m-E_n)]t} + r\frac{1-e^{-[r+i(E_m-E_n)]t}}{r+i(E_m-E_n)} \right\} \ket{\psi_m} \bra{\psi_n},
\end{equation}
where $E_n = -\omega n$. Its fidelity $\mathcal{F}_r(t):=\bra{\psi(0)}\hat\rho_r(t)\ket{\psi(0)}$ reads then:
\begin{align}
&\mathcal{F}_r(t)= \sum_{m,n=0}^{L} |c_m|^2 |c_n|^2 \left\{e^{-[r+i(E_m-E_n)]t} + r\frac{1-e^{-[r+i(E_m-E_n)]t}}{r+i(E_m-E_n)} \right\},
\end{align}
with $c_n=\braket{n|\psi_r}$. In the long-time limit one gets the stationary state:
\begin{align}
\label{S:Eq:fidelity_reset}
\lim_{t\rightarrow +\infty} \mathcal{F}_r(t)&=\sum_{m,n=0}^{L} |c_m|^2 |c_n|^2 \frac{r}{r-i\omega(m-n)}=\\
& = \frac{1}{2}\sum_{m,n=0}^{L} |c_m|^2 |c_n|^2\left[  \frac{r}{r-i\omega(m-n)}+ \frac{r}{r+i\omega(m-n)} \right]= \sum_{m,n=0}^{L} |c_m|^2 |c_n|^2 \frac{r^2}{r^2+\omega^2 (m-n)^2}.
\end{align}

We now specialize to the case where the initial state $\ket{\psi(0)}$ takes the coherent-state form:
\begin{equation}
    \ket{\alpha}=\frac{1}{(1+|\alpha|^2)^{L/2}}e^{\alpha \hat Q^{\dag}}\ket{0},
\end{equation}
as in Eq.~\eqref{Eq:coh_state} of the main text. The fidelity is then obtained from Eq.\eqref{S:Eq:fidelity_reset} by plugging in the expression of the expansion coefficients of the state $\ket{\alpha}$ over the scarred eigenstates:
\begin{equation}
\label{S:coefficients}
    c_n(t)=\braket{\psi_n|\alpha(t)}= \frac{\sqrt{\binom{L}{n}} \alpha^n(t)}{(1+|\alpha|^2)^{L/2}}, \quad
    |c_n|^2=|\braket{\psi_n|\alpha}|^2 = \frac{\binom{L}{n} |\alpha|^{2n}}{(1+|\alpha|^2)^{L}},
\end{equation}
with the time evolution of the complex parameter $\alpha(t)$ given in Eq.~\eqref{eq:coh_state_evolution}. The stationary value of the fidelity in presence of resetting reads for such a specific choice of the initial state reads finally:
\begin{equation} \label{S:Eq:fidelity_reset_coh}
\mathcal{F}_r(+\infty) = \frac{1}{{(1+|\alpha|^2)^{2L}}}\sum_{m,n=0}^{L} \binom{L}{m}\binom{L}{n} |\alpha|^{2(m+n)} \frac{r^2}{r^2+\omega^2 (m-n)^2}.
\end{equation}
We notice that Eq.~\eqref{S:Eq:fidelity_reset_coh} equals the expression for the purity of the density matrix in the presence of resetting $\mathcal{F}_r(+\infty)=\mbox{Tr}[\hat\rho_r^2 (+\infty)]$. To prove such a relation, we express $\mbox{Tr}[\hat\rho_r^2 (+\infty)]$ as:
\begin{equation}\label{S:Eq:proof_fid_2ren}
    \mbox{Tr}[\hat\rho_r^2 (+\infty)]= r^2\int_0^{+\infty}d\tau_1 \int_0^{+\infty} d\tau_2 e^{-r(\tau_1 +\tau_2)} |\braket{\alpha(\tau_1)|\alpha(\tau_2)}|^2.
\end{equation}
After the change of variable $\tau_1'=\tau_1$, $\tau_2'=\tau_2-\tau_1$ and the observation that $ |\braket{\alpha(\tau_1)|\alpha(\tau_2)}|^2 =  |\braket{\alpha(0)|\alpha(\tau_2-\tau_1)}|^2$, Eq.~\eqref{S:Eq:proof_fid_2ren} can be rewritten as:
\begin{equation}\label{S:Eq:proof_long}
\begin{split}
    &r^2\int_0^{+\infty}d\tau_1 \int_0^{+\infty} d\tau_2\, e^{-r(\tau_1 +\tau_2)} |\braket{\alpha(\tau_1)|\alpha(\tau_2)}|^2 =\\
    &=r^2\int_0^{+\infty}d\tau_2' \int_0^{+\infty} d\tau_1'\, e^{-r(2\tau_1' +\tau_2')} |\braket{\alpha(0)|\alpha(\tau_2')}|^2 + r^2\int_{-\infty}^{0}d\tau_2' \int_{-\tau_2'}^{+\infty} d\tau_1' \, e^{-r(2\tau_1' +\tau_2')} |\braket{\alpha(0)|\alpha(\tau_2')}|^2 =\\
    &=\frac{r}{2}\int_0^{+\infty} d\tau_2'\, e^{-r\tau_2'} \braket{\alpha(0)|\alpha(\tau_2')}|^2 +\frac{r}{2}\int_{-\infty}^{0} d\tau_2'\, e^{r\tau_2'} \braket{\alpha(0)|\alpha(\tau_2')}|^2.
\end{split}
\end{equation}
By eventually performing a last change of variable $\tau_2''=-\tau_2'$ in the last integral of Eq.~\eqref{S:Eq:proof_long}, after noticing that $ |\braket{\alpha(0)|\alpha(\tau_2)}|^2= |\braket{\alpha(\tau_2)|\alpha(0)}|^2= |\braket{\alpha(0)|\alpha(-\tau_2)}|^2$, one concludes that, indeed, $\mathcal{F}_r(+\infty)=\mbox{Tr}[\hat\rho_r^2 (+\infty)]$. Furthermore, drawing on the results presented in that section, we conclude that the sparse-resetting $r\rightarrow 0^+$ discussed in the main text can be estimated by restricting in Eq.~\eqref{S:Eq:fidelity_reset_coh} the sum only to the terms with $m=n$; the latter, in the large-size limit, has the asymptotic scaling $L^{-1/2}$ that generalizes the case $\alpha=1$ highlighted in the main text (see also Eq.~\eqref{S:E1} below).  

The relation in Eq.\eqref{S:Eq:fidelity_reset_coh} can be further simplified in the case $\alpha = 1$: indeed, the latter can be manipulated by splitting the different contributions to it according to the value of $d=|m-n|=0,\,\dots,\, L$, as follows:
\begin{equation}
    \begin{split}
        &\frac{1}{2^{2L}}\sum_{m=0}^L\sum_{n=0}^L \binom{L}{m}\binom{L}{n}\frac{r^2}{r^2+\omega^2 (m-n)^2}=\\
        &=\frac{1}{2^{2L}}\sum_{d=1}^L \frac{r^2}{r^2+\omega^2 d^2} \left[\sum_{m=0}^{L-d}\binom{L}{m}\binom{L}{m+d}+\sum_{m=d}^{L}\binom{L}{m}\binom{L}{m-d}\right]+\frac{1}{2^{2L}}\sum_{m=0}^L \binom{L}{m}^2.
    \end{split}
\end{equation}
By the change of variable $m'=m-d$ in the second summation over $m$ and the use of the property of the binomial $\binom{L}{k}=\binom{L}{L-k}$, one can rewrite the above expression as:
\begin{equation}
    \begin{split}
       &\frac{1}{2^{2L}}\sum_{d=1}^L \frac{r^2}{r^2+\omega^2 d^2} \left[2\sum_{m=0}^{L-d}\binom{L}{m}\binom{L}{L-d-m}\right]+\frac{1}{2^{2L}}\binom{2L}{L}=\frac{2}{2^{2L}}\sum_{d=1}^L \binom{2L}{L-d}\frac{r^2}{r^2+\omega^2 d^2} +\frac{1}{2^{2L}}\binom{2L}{L}=\\
       &=\frac{2}{2^{2L}}\sum_{d=0}^{L-1} \binom{2L}{d}\frac{r^2}{r^2+\omega^2 (L-d)^2}+\frac{1}{2^{2L}}\binom{2L}{L}, 
    \end{split}
\end{equation}
where we used the property $\sum_{m=0}^{L-d}\binom{L}{m}\binom{L}{L-d-m}=\binom{2L}{L-d}$ and, in the last step, we performed the change of variable $d'=L-d$. Finally, keeping in mind the factor $2$ and observing that, after the change of variable $d'=2L-d$, the summation over $d$ can be rewritten as:
\begin{equation}
  \sum_{d=0}^{L-1} \binom{2L}{d}\frac{r^2}{r^2+\omega^2 (L-d)^2}=\sum_{d=L+1}^{2L} \binom{2L}{2L-d}\frac{r^2}{r^2+\omega^2 (d-L)^2}=\sum_{d=L+1}^{2L} \binom{2L}{d}\frac{r^2}{r^2+\omega^2 (d-L)^2},
\end{equation}
one can rewrite Eq.~\eqref{S:Eq:fidelity_reset_coh} in the case $\alpha=1$ as:
\begin{equation}
 \mathcal{F}_r(+\infty)= \frac{1}{2^{2L}}\sum_{d=0}^{2L} \binom{2L}{d}\frac{r^2}{r^2+\omega^2 (L-d)^2}.
\end{equation}
This result matches Eq.~\eqref{eq:fidelity_stationary_value} of the main text. 

\section{Local observables}
\label{S:local_obsevables}
In this Appendix we provide the derivation for Eqs.~\eqref{Eq:observable_vs_t} and \eqref{Eq:cc} in Sec.~\ref{sec:correlations} for the dynamics and stationary values of local observables in the presence of resetting. In addition, we also report the derivation of Eq.~\eqref{eq:long_range_correlations} for the correlation functions of the quasi-particle creation operator in the stationary state induced by resetting. 

First, we observe that the operators that are diagonal in the scarred subspace, i.e., that satisfy $\bra{\psi_m}\hat O\ket{\psi_n}\propto \delta_{m,n}$, have time-independent expectation values at all times, as:
\begin{equation}
\mbox{Tr}[\hat O \hat\rho_r(t)]=\sum_{l=0}^L \bra{\psi_l} \hat O \hat\rho_r(t) \ket{\psi_l} = \sum_{l=0}^L \bra{\psi_l} \hat O \ket{\psi_l}\bra{\psi_l} \hat\rho_r(t) \ket{\psi_l}=\sum_{l=0}^L |c_l|^2\bra{\psi_l} \hat O \ket{\psi_l},
\label{eq:diagonal_observable}
\end{equation}
where we assumed that the initial and reset states are equal and of the form in Eq.~\eqref{eq:scars_initial}. This ensures that the dynamics remains restricted to the scar subspace at all times. We also exploited the spectral decomposition of the time-evolved density matrix $\hat{\rho}_r(t)$ in Eq.~\eqref{S:Eq:generic_rho_r}.
We remark that the property in Eq.~\eqref{eq:diagonal_observable} applies as well for purely unitary evolution. 

Now, we consider the operator $\hat O_j = e^{-ikj} \hat q_j $, with $k=\pi$ and $\hat q_j \in\left\{ \hat c_{j,\downarrow} \hat c_{j,\uparrow}, \frac{\left( \hat S^-_j \right)^2}{2} \right\}$, which is instead off-diagonal in the scarred subspace and features infinitely long-lived oscillations in time when unitarily evolving from states with significant overlap with scarred subspace. Moreover, we fix, as usual, the initial state to be $\ket{\alpha}\bra{\alpha}$. We compute its expectation value in the presence of a finite resetting rate as follows:
\begin{align}
& \mbox{Tr}[\hat O_j \hat\rho_r(t)]= e^{-rt} \bra{\alpha(t)} \hat O_j \ket{\alpha(t)} + r\int_0^t d\tau\, e^{-r\tau} \bra{\alpha(\tau)} \hat O_j \ket{\alpha(\tau)}.
\end{align}
Since
\begin{equation}
    \bra{\alpha} \hat O_j \ket{\alpha} = \frac{\alpha}{1+|\alpha|^2}, 
\end{equation}
the final result reads
\begin{equation}
  \mbox{Tr}[\hat O_j \hat\rho_r(t)]= e^{-rt +i\omega t}\frac{\alpha}{1+|\alpha|^2}  + \frac{r}{r-i\omega}\frac{\alpha}{1+|\alpha|^2}\left( 1- e^{-rt+i\omega t} \right).
\label{S:O_j_time}
\end{equation}
The stationary value is simply given by
\begin{equation}
  \mbox{Tr}[\hat O_j \hat\rho_r(+\infty)]=  \frac{r}{r-i\omega}\frac{\alpha}{1+|\alpha|^2}.
\label{S:O_j_stationary}
\end{equation}
As in the case of fidelity, we thus observe that the presence of resetting destroys the characteristic phenomenology of quantum many-body scars by preventing revivals and driving the system to a stationary state. Equations \eqref{S:O_j_time} and \eqref{S:O_j_stationary} match Eqs.~\eqref{Eq:observable_vs_t} and \eqref{Eq:cc} of the main text.

We end this section by evaluating the connected correlation function of the quasiparticle ladder operator $\hat Q^{\dag}$ linked to the class of towers of scars studied in this manuscript. The stationary state connected correlation function for this operator is given by
\begin{equation} \label{S:Eq:conn_corr_Q}
    \mbox{Tr}\left[\hat Q \hat Q^{\dag} \hat\rho_r(+\infty)\right]-\mbox{Tr}\left[\hat Q \hat\rho_r(+\infty)\right]\mbox{Tr}\left[ \hat Q^{\dag} \hat\rho_r(+\infty)\right] = \mbox{Tr}\left[\hat Q \hat Q^{\dag} \hat\rho_r(+\infty)\right]-\Biggl|\mbox{Tr}\left[ \hat Q^{\dag} \hat\rho_r(+\infty)\right]\Biggr|^2.
\end{equation}
The first term on the right hand side of the equation is the expectation value of a diagonal operator ($\hat{Q}\hat{Q}^{\dagger}$) in the scarred subspace and it is consequently time-independent, as discussed above in Eq.~\eqref{eq:diagonal_ensemble}. Its expectation value over the initial state is given by
\begin{equation}
\label{S:connected}
    \mbox{Tr}\left[\hat Q \hat Q^{\dag} \hat\rho_r(+\infty)\right] = \bra{\alpha}\hat Q \hat Q^{\dag} \ket{\alpha} = L(L-1)\frac{|\alpha|^2}{(1+|\alpha|^2)^2}+L\frac{1}{1+|\alpha|^2}.
\end{equation}
On the other hand, the second term on the right hand side of Eq.~\eqref{S:Eq:conn_corr_Q} is the expectation value of a purely off-diagonal operator in the scarred subspace. Its stationary state expectation value is
\begin{equation}
\label{S:disconnected}
  \mbox{Tr}\left[ \hat Q^{\dag} \hat\rho_r(+\infty)\right]= r\int_0^{+\infty} d\tau \, e^{-r\tau} \bra{\alpha(\tau)} \hat Q^{\dag} \ket{\alpha(\tau)} = L \frac{\alpha^*}{1+|\alpha|^2} \frac{r}{r+i\omega}, 
\end{equation}
where we employed the result $\bra{\alpha(\tau)} \hat Q^{\dag} \ket{\alpha(\tau)}=L \alpha^* e^{-i\omega\tau}/(1+|\alpha|^2)$ and Eq.~\eqref{eq:coh_state_evolution} for the time-evolution of a scar coherent state. Putting together Eqs.~\eqref{S:connected} and \eqref{S:disconnected}, one eventually obtains the connected correlation function. In the thermodynamic limit this is given by
\begin{equation}
    \lim_{L\rightarrow +\infty} \frac{1}{L^2} \left\{ \mbox{Tr}\left[\hat Q \hat Q^{\dag} \hat\rho_r(+\infty)\right]-\Biggl|\mbox{Tr}\left[ \hat Q^{\dag} \hat\rho_r(+\infty)\right]\Biggr|^2\right\} = \frac{|\alpha|^2}{(1+|\alpha|^2)^2} - \frac{r^2
    }{r^2 +\omega^2}\frac{|\alpha|^2}{(1+|\alpha|^2)^2}=\frac{\omega^2
    }{r^2 +\omega^2}\frac{|\alpha|^2}{(1+|\alpha|^2)^2}.
\end{equation}
This result matches Eq.~\eqref{eq:ODLR_scar_reset} of the main text and it shows that the stationary state obtained via resetting displays off-diagonal long-range order since it can be understood as a condensate of $\pi$-momentum quasiparticles.

\section{Dynamical correlation function}\label{S:dyn_corr_fun}
In this Appendix we derive Eqs.~\eqref{Eq:C_r_tt} and \eqref{eq:correlation_function_Q_dyn_final} of the main text. 

Let us proceed to evaluate the two-point correlation function $C_{r,\hat Q^{\dag}}(t_1,t_2)$ of the raising operator $\hat Q^{\dag}$ over a coherent state $\ket{\alpha}$; we start by evaluating its reset-free version, namely:
\begin{equation}
    C_{\hat Q^{\dag}}(t_1,t_2) =  \bra{\alpha}\hat Q(t_2) \hat Q^{\dag}(t_1)\ket{\alpha}-\bra{\alpha}\hat Q(t_2) \ket{\alpha}\bra{\alpha} \hat Q^{\dag}(t_1)\ket{\alpha}.
\end{equation}
The disconnected term is time-translation-invariant, since:
\begin{equation}
\label{S:Eq:disconnected}
    \bra{\alpha}\hat Q^{\dag}(t_1)\ket{\alpha}= L \frac{\alpha^* e^{-i\omega t_1}}{1+|\alpha|^2} \implies \bra{\alpha}\hat Q(t_2) \ket{\alpha}\bra{\alpha} \hat Q^{\dag}(t_1)\ket{\alpha} = L^2 \frac{|\alpha|^2}{(1+|\alpha|^2)^2} e^{i\omega(t_2 - t_1)},
\end{equation}
while the connected contribution, which turns out to depend on $t_2 - t_1$ only as well, can be evaluated as follows:
\begin{equation}
\label{S:Eq:two_pt_step}
    \bra{\alpha}\hat Q(t_2) \hat Q^{\dag}(t_1)\ket{\alpha}= \sum_{m,n=0}^L c_{m}^*(t_2) c_n (t_1) \bra{\psi_m} \hat Q e^{-i\hat H (t_2 - t_1)} \hat Q^{\dag} \ket{\psi_n}.
\end{equation}
Here we have expanded the coherent state in the scarred-state basis with coefficients in Eq.~\eqref{S:coefficients}, with $\alpha(t)= \alpha e^{i\omega t}$ from Eqs.~\eqref{Eq:coh_state} and \eqref{eq:coh_state_evolution}. Noticing that in Eq.~\eqref{S:Eq:two_pt_step} one has $\hat Q^{\dag}\ket{\psi_n}\propto \ket{\psi_{n+1}}$ and $\bra{\psi_m}\hat Q \propto \bra{\psi_{m+1}}$, one can rewrite it as follows
\begin{align}
   \bra{\alpha}\hat Q(t_2) \hat Q^{\dag}(t_1)\ket{\alpha}&= \sum_{m=0}^{L-1} c_{m}^*(t_2) c_m (t_1) e^{i\omega(m+1) (t_2 - t_1)}  \bra{\psi_m} \hat Q \hat Q^{\dag} \ket{\psi_m} = \nonumber \\
   &=e^{i\omega (t_2 -t_1)} \sum_{m=0}^{L-1} |c_m(0)|^2 \bra{\psi_m} \hat Q \hat Q^{\dag} \ket{\psi_m} = e^{i\omega(t_2 -t_1)} \bra{\alpha} \hat Q \hat Q^{\dag} \ket{\alpha}.
\label{S:Eq:connected}   
\end{align}
The expression of the correlation function $\braket{\alpha|\hat{Q}(t_2)\hat{Q}^{\dagger}(t_1)|\alpha}$ can be obtained following steps analogous to those carried for Eq.~\eqref{S:connected} and it leads to
\begin{equation}
   \bra{\alpha} \hat Q \hat Q^{\dag} \ket{\alpha} = L(L-1)\frac{|\alpha|^2}{(1+|\alpha|^2)^2} + L\frac{1}{1+|\alpha|^2}.
\end{equation}
Therefore, collecting all the terms in Eqs.~\eqref{S:Eq:disconnected} and \eqref{S:Eq:connected} we obtain the final result
\begin{equation}
    C_{r,\hat Q^{\dag}}(t_1,t_2) = e^{-r(t_2 - t_1)} C_{\hat Q^{\dag}}(t_1,t_2) =  e^{-r(t_2 - t_1) +i\omega(t_2-t_1)} L \frac{1}{(1+|\alpha|^2)^2},
\label{S:final_dyn_corr}
\end{equation}
which matches \eqref{eq:correlation_function_Q_dyn_final} of the main text. We notice that the reset-free correlation function $C_{\hat{Q}^{\dagger}}(t_1,t_2)$ is a function of the time difference $t_2-t_1$ in spite of the state $\ket{\alpha}$ not being an eigenstate of the Hamiltonian. The latter is, indeed, a sufficient but not necessary condition for the dynamical correlation function to be time-translation invariant. 

For the sake of completeness, we can repeat the calculation for a different initial state, which we set to be a fixed scar eigenstate $\ket{\psi_n}$. In such case, the reset-free two-point correlation function is solely given by the connected term and thus one immedetialy obtains
\begin{equation}\label{Eq:corr_ex}
    C_{\hat Q^{\dag}}(t_1,t_2)= e^{i\omega(t_2-t_1)}(L-n)(n+1).
\end{equation}
Plugging Eq.\eqref{Eq:corr_ex} into Eq.\eqref{S:final_dyn_corr}, one obtains the result:
\begin{equation}
    C_{r,\hat Q^{\dag}}(t_1,t_2)=(L-n)(n+1)e^{-(r-i\omega)(t_2-t_1)}.
\end{equation}
We therefore conclude that stochastic resetting generically destroys long-range time order characterizing quantum scars by introducing an exponential damping, as a function of time, of the correlations.

\section{Rényi entropies after a quench from a coherent state}
\label{S:entropies}
In this Appendix we provide the calculations leading to all the results of Sec.~\ref{sec:entropy_mix}. In Subsec.~\ref{App:steady_S2}, we obtain the  exact equation \eqref{Eq:2_Renyi} for the purity of the stationary state and Rényie-2 entropy in the presence of resetting. In Subsec.~\ref{app:long_L_2renyie} we get the large $L$ asymtptotics of this result by saddle-point analysis. In Subsec.~\ref{app:long_L_n_renyie}, we show that the same asymptotic scaling at large $L$ holds for all the Rényi entropies ($n>2$). In Subsec.~\ref{app:time_dependence_Renyie}, we derive Eq.~\eqref{app:time_dependence_Renyie}, which is the main result that we used in Fig.~\ref{Fig:4} to compute the time dynamics of the Rényi-2 entropy. In Subsec.~\ref{app:rsmall_reduced_rho}, we show that the NESS obtained from resetting for small resetting $r$ has the structure in Eq.~\eqref{eq:reduced_state_scar} analogous to the reduced density matrix of a scarred eigenstate.

\subsection{Derivation of the stationary state expression of $2$-Rényi entropy}
\label{App:steady_S2}

In order to quantify the degree of mixedness in the stationary state of the dynamics, we introduce the $n$-th Rényi entropy of a state $\hat \rho$ through the definition:
\begin{equation}
\begin{split}
S_n(\hat\rho)=\frac{1}{1-n}\log \left(\mbox{Tr}[\hat \rho^n]\right).
\end{split}
\end{equation}
In absence of resetting, the time evolution amounts to the replacement $\alpha\rightarrow \alpha(t)=\alpha e^{i\omega t}$. The state remains pure at all times and thus does not develop mixedness. In presence of resetting, instead, the stationary state writes:
\begin{align}
\hat \rho_r(+\infty)= r\int_0^{+\infty} d\tau\, e^{-r\tau} \ket{\alpha(\tau)}\bra{\alpha(\tau)}.
\end{align}
The $n$-th Rényi entropy can be evaluated by taking the trace of $\hat\rho_r^n$, which reads:
\begin{align}
\hat\rho^n_{r}(+\infty) = r^n \left(\prod_{j=1}^n \int_0^{+\infty} d\tau_j e^{-r\tau_j}\right)\left(\prod_{l=1}^{n-1}\bra{\alpha(\tau_l)}\ket{\alpha(\tau_{l+1})}\right) \ket{\alpha(\tau_1)}\bra{\alpha(\tau_n)},
\end{align}
where we denoted $\ket{\alpha(\tau)}:=\ket{\alpha e^{i\omega\tau}}$.
Using the results:
\begin{equation}
\phi(\tau_{l+1}-\tau_l):= \braket{\alpha(\tau_l)|\alpha(\tau_{l+1})}= (1/(1+|\alpha|^2)^L)(1+|\alpha|^2 e^{i\omega (\tau_{l+1}-\tau_l)})^{L}=\mbox{Tr}[\ket{\alpha(\tau_{l+1})}\bra{\alpha(\tau_{l})}],
\end{equation}
one obtains the following result for the exponentiated $n$-th Rényi entropy:
\begin{equation} \label{App:n_renyi}
\begin{split}
&\mbox{Tr}[\hat\rho^n_{r}(+\infty)] = r^n \left(\prod_{j=1}^n \int_0^{+\infty} d\tau_j e^{-r\tau_j}\right)\left(\prod_{l=1}^{n-1}\phi (\tau_{l+1}-\tau_l)\right) \phi(\tau_1 - \tau_n)=\\
&= |\mathcal{N}_{L}|^{2n} r^n \sum_{l_1,\dots, l_n=0}^{L}\left(\prod_{j=1}^n \binom{L}{l_j} |\alpha|^{2 l_j} \right)\left(\prod_{k=1}^n \frac{1}{r-i\omega(l_{k-1}-l_k)} \right).
\end{split}
\end{equation}
In the last product of the above expression we imply the identification $l_0=l_n$. In the case $n=2$, e.g., the full expression reduces to:
\begin{equation}\label{Eq:renyi2}
\begin{split}
&\mbox{Tr}[\hat\rho^2_{r}(+\infty)] = \frac{1}{(1+|\alpha|^2)^{2L}}\sum_{m,n=0}^{L} \binom{L}{m}\binom{L}{n} |\alpha|^{2(m+n)}\frac{r^2}{r^2+\omega^2(m-n)^2}.
\end{split}
\end{equation}
This equation reproduces Eq.~\eqref{Eq:2_Renyi} of the main text.

\subsection{Stationary-state 2-Rényi entropy for weak and finite resetting rate}
\label{app:long_L_2renyie}

We now study analytically the asymptotic behavior of the Rényi-$2$ entropy in Eq.~\eqref{Eq:renyi2} for large $L$. We do this first in the weak resetting limit $r\rightarrow 0^+$. Then, we extend the result to generic values of the resetting rate $r$. In both the cases we find logarithmic scaling of the mixedness as in Eq.~\eqref{eq:entropy_asymptotic_scaling} of the main text. 

In the weak resetting limit, only the terms with $m=n$ in the summation of Eq.~\eqref{Eq:renyi2} will not contain a factor of $r^2$ in front, so that we approximate the sum by keeping only the diagonal $m=n$ terms
\begin{equation}\label{Eq:diag_2}
\mbox{Tr}[\hat\rho^2_{r}(+\infty)] \approx \sum_{n=0}^L |c_n|^4=\frac{1}{(1+|\alpha|^2)^{2L}}\sum_{n=0}^{L} \binom{L}{n}^2 |\alpha|^{4n}.
\end{equation}
Then, we further consider the thermodynamic limit $n,L\gg 1$, with $n/L \sim \mathcal{O}(1)$ so that the density of quasiparticles remains finite. We then approximate the binomial coefficients \eqref{S:coefficients} with the help of Stirling's formula $n!\sim \sqrt{2\pi n} n^n e^{-n}$ and we introduce the continuous variable $y=n/L\in [0,1]$. In this way one rewrites the sum \eqref{Eq:diag_2} as
\begin{equation}
\begin{split}
&\frac{1}{(1+|\alpha|^2)^{2L}}\sum_{n=0}^{L} \binom{L}{n}^2 |\alpha|^{4n} \approx \frac{1}{2\pi (1+|\alpha|^2)^{2L}}\int_0^1 dy \frac{1}{y(1-y)}e^{-2Lf(y)},
\label{eq:integral_saddle}
\end{split}
\end{equation}
where 
\begin{equation}
f(y)=-2\log(|\alpha|)y+y\log y +(1-y)\log(1-y). 
\label{eq:saddle_point_eq}
\end{equation}
In the thermodynamic limit the integral in Eq.~\eqref{eq:integral_saddle} can be evaluated by saddle point method. The integral is therefore dominated by the behavior of the integrand close to the minimum of $f(y)$, which is given by
\begin{equation}
y^*=\frac{|\alpha|^2}{1+|\alpha|^2}.
\label{S:y_star_saddle}
\end{equation}
In this way we obtain the final result:
\begin{equation}
\label{S:Eq:asympt_fid}
\begin{split}
&\frac{1}{2\pi (1+|\alpha|^2)^{2L}}\int_0^1 dy \frac{1}{y(1-y)}e^{-2Lf(y)}\approx \frac{1}{2\pi (1+|\alpha|^2)^{2L} \,y^*(1-y^*)} e^{-2L f(y^*)} \sqrt{\frac{\pi}{L f''(y^*)}} = \frac{1+|\alpha|^2}{2|\alpha| \sqrt{\pi L}}.
\end{split}
\end{equation}
The second Rényi entropy is thus approximated as:
\begin{equation} \label{S:Eq:small_r_estimate}
S_2(\hat{\rho}_r(+\infty))  -\log \mbox{Tr}[\hat\rho^2_{r}(+\infty)]\approx   \log\left[ \frac{2|\alpha| \sqrt{\pi L}}{1+|\alpha|^2}\right],
\end{equation}
which reproduces the logarithmic scaling reported in Eq.~\eqref{eq:entropy_asymptotic_scaling} of the main text. Let us now try to give a rigorous justification of the diagonal approximation. The $2$-Rényi entropy from Eq.~\eqref{Eq:renyi2} can be rewritten as:
\begin{align}
 -\log \mbox{Tr}[\hat\rho^2_{r,A}(+\infty)]=&-\log\left[\frac{1}{(1+|\alpha|^2)^{2L}}\sum_{n=0}^{L} \binom{L}{n}^2 |\alpha|^{4n}\right. + \nonumber \\
 &\quad \quad \, \,\,\,\left.+\frac{1}{(1+|\alpha|^2)^{2L}}\sum_{m=0}^{L}\sum_{n=0,\, n \neq m}^{L} \binom{L}{m}\binom{L}{n} |\alpha|^{2(m+n)}\frac{r^2}{r^2+\omega^2(m-n)^2}\right].  
\end{align}
Noticing that the first term does not depend on $r$ and that:
\begin{align}\label{Eq:rig_bound}
\frac{r^2}{r^2+\omega^2(m-n)^2}\leq \frac{r^2}{r^2+\omega^2},
\end{align}
for the allowed values of $m\neq n$, one gets the bounds:
\begin{align}\label{Eq:saddle_renyi2}
-\log\left[\frac{1}{(1+|\alpha|^2)^{2L}}\sum_{n=0}^{L} \binom{L}{n}^2 |\alpha|^{4n} + \frac{r^2}{r^2+\frac{\omega^2}{\hbar^2}}\right]\leq -\log \mbox{Tr}[\hat\rho^2_{r}(+\infty)]\leq -\log\left[\frac{1}{(1+|\alpha|^2)^{2L}}\sum_{n=0}^{L} \binom{L}{n}^2 |\alpha|^{4n} \right].
\end{align}
We used the fact that the function $-\log x$ is monotonically decreasing. Therefore, after taking the $r\rightarrow 0^+$ limit, the value of the $2$-Rényi entropy is well defined and given by Eq.~\eqref{Eq:diag_2}.

We might now ask ourselves whether we can show if the logarithmic scaling of the $2$-Rényi entropy will occur regardless of the magnitude of the resetting rate, as long as the size of the system is sufficiently large. To this end, we start from Eq.~\eqref{Eq:renyi2} and consider the large-size limit without explicitly neglecting the $m\neq n$ terms of the summation that gives the $2$-Rényi entropy. In terms of the rescaled variables $x=m/L$, $y=n/L$, we obtain the representation:
\begin{align}\label{Eq:univ}
&\mbox{Tr}[\hat\rho^2_{r}(+\infty)] \approx\frac{L}{2\pi(1+|\alpha|^2)^{2L}} \int_0^1 dx \int_0^1 dy \frac{e^{-L[f(x) + f(y)]}}{\sqrt{x(1-x)y(1-y)}} \frac{1}{1+\left(\frac{\omega L}{r} \right)^2 (x-y)^2},
\end{align}
where $f(x)$ has the expression in Eq.~\eqref{eq:saddle_point_eq}. Introducing the notation $\gamma=\omega L/r$ and exploiting the Fourier representation:
\begin{equation}
    \frac{1}{1+\gamma^2 (x-y)^2}=\frac{1}{2\gamma}\int_{-\infty}^{+\infty} dk e^{ik(x-y)-|k|/\gamma},
\end{equation}
one may rewrite Eq.~\eqref{Eq:univ} as:
\begin{equation}
    \begin{split}
        \mbox{Tr}[\hat\rho^2_{r}(+\infty)] \approx \frac{L}{4\pi\gamma(1+|\alpha|)^{2L}}\int_{-\infty}^{+\infty} dk\, e^{-|k|/\gamma}\int_0^1 dx \int_0^1 dy \frac{e^{-Lf(x)+ik(x-x^*)}e^{-Lf(y)-ik(y-y^*)}}{\sqrt{x(1-x)y(1-y)}},
    \end{split}
\end{equation}
where $x^*=y^*=\frac{|\alpha|^2}{1+|\alpha|^2}$ is the coordinate of the minimum of the function $f(x)$ introduced before. After expanding the functions $f(x)$ and $f(y)$ to quadratic order and evaluating the resulting integrals in the $x$ and $y$ variables via the saddle-point method, several algebraic simplifications lead to the expression:
\begin{equation}
    \mbox{Tr}[\hat\rho^2_{r}(+\infty)] \approx \frac{1}{2}\int dx \, e^{-|x|-\frac{\omega^2 L x^2}{r^2 f''(x^*)}},  
\end{equation}
where $f''(x^*)=\frac{(1+|\alpha|^2)^2}{|\alpha|^2}$. A further saddle-point estimate gives the result:
\begin{equation}
    \mbox{Tr}[\hat\rho^2_{r}(+\infty)] \sim \frac{r(1+|\alpha|^2)}{2\omega |\alpha|}\sqrt{\frac{\pi}{L}},  
\end{equation}
in the large-size limit, thus recovering the same logarithmic scaling \eqref{S:Eq:small_r_estimate} obtained in the sparse-resetting limit. We therefore conclude that logarithmic scaling of the mixedness in the stationary state applies for generic values of the resetting rate $r$. The value of the resetting $r$ determines, however, the onset of such asymptotic scaling, which takes places at larger values of $L$ as the resetting rate is increased (see Fig.~\ref{Fig:entropy_scaling} of the main text).  


\subsection{Small-$r$ limit for the stationary $n$-Rényi entropy}
\label{app:long_L_n_renyie}
In the section, we show that logarithmic scaling is a generic properties of Rényie entropies ($n>2$) of the stationary state and it is not restricted to Rényie-$2$ entropy only. In this case, we can perform the analytical calculation for the weak-resetting limit only. We, however, numerically checked that the same logarithmic scaling holds also for finite values of the resetting rate.

When the resetting rate is sufficiently small, we can approximate Eq.~\eqref{Eq:n_Rényi_coh_states} with its diagonal terms satisfying $l_j=l_k \,\,\forall j,k$, and then apply a large-$L$ saddle-point estimate of the resulting expression, as done for the $n=2$ case. In formulas:
\begin{equation}
\begin{split}
&\mbox{Tr}[\hat\rho^n_{r}(+\infty)] = \frac{1}{(1+|\alpha|^2)^{nL}}\sum_{m=0}^{L} \binom{L}{m}^n |\alpha|^{2nm}\approx \frac{L}{(1+|\alpha|^2)^{nL}}\int_0^1 dx \frac{e^{-nLf(x)}}{[2\pi L x(1-x)]^{n/2}},
\end{split}
\end{equation}
where we introduced the rescaled continuous variable $x=m/L\in [0,1]$ and the function $f(x)$ has the same structure as in Eq.~\eqref{eq:saddle_point_eq}. The saddle-point result then reads
\begin{equation}
\begin{split}
&S_n(\hat\rho_r(+\infty))=\frac{1}{1-n}\log \mbox{Tr}[\hat\rho^n_{r}(+\infty)] \approx \frac{1}{1-n}\log \left\{ \frac{1}{\sqrt{n}} \left[\frac{2\pi |\alpha|^2 L}{(1+|\alpha|^2)^2}\right]^{\frac{1-n}{2}}   \right\},
\end{split}
\end{equation}
which, for generic $n$, gives the large-size scaling:
\begin{equation}
S_n(\hat{\rho}_r(+\infty))\sim \frac{1}{2}\log L.   
\end{equation}
This asymptotic form coincides with Eq.~\eqref{S:Eq:small_r_estimate} derived earlier for the second Rényi entropy.

\subsection{The sparse-resetting limit of the stationary density matrix as a reduced density matrix}
\label{app:rsmall_reduced_rho}
In this section, we show that the reduced half-chain density matrix associated to a generic scarred eigenstate $\ket{\psi_n}$ takes the form in Eq.~\eqref{eq:reduced_state_scar}, which is analogous to the diagonal ensemble in Eq.~\eqref{eq:diagonal_ensemble}. This is the reason why the mixedness of the stationary state obtained via resetting shows the same scaling, as a function of the system size, as the entanglement entropy of a scarred eigenstate. 

The half-chain reduced density matrix of a scarred eigenstate $\ket{\psi_n}$ can be computed from the following rewriting  of the state $\ket{\psi_{n,L}}$ in Eq.~\eqref{S:Eq:eta_pair_tower} (we report the system size $L$ explicitly as a subscript here)
\begin{equation}
\label{S:Eq:SD_scars}
\ket{\psi_{n,L}}=\frac{1}{\sqrt{\binom{L}{n}}}\sum_{1\leq j_1<\dots<j_n\leq L}e^{ik(\sum_{l=1}^n j_l)} q^{\dag}_{j_1} \dots q^{\dag}_{j_n}\ket{0}=\sum_{m=\mathrm{max}(0,n-L/2)}^{\mathrm{min}(n,L/2)} \sqrt{\frac{\binom{L/2}{m}\binom{L/2}{n-m}}{\binom{L}{n}}} \ket{\psi_{m,L/2}^{(l)}}\otimes \ket{\psi_{n-m,L/2}^{(r)}},
\end{equation}
Here we introduced the notation:
\begin{equation}
\begin{split}
& \ket{\psi_{p,L/2}^{(l)}}=\frac{1}{\sqrt{\binom{L/2}{p}}}\sum_{1\leq j_1<\dots<j_n\leq \frac{L}{2}}e^{ik(\sum_{l=1}^p j_l)} \hat q^{\dag}_{j_1} \dots \hat q^{\dag}_{j_p}\ket{0},\\
&\ket{\psi_{p,L/2}^{(r)}}=\frac{1}{\sqrt{\binom{L/2}{p}}}\sum_{\frac{L}{2}+1\leq j_1<\dots<j_n\leq L} e^{ik(\sum_{l=1}^p j_l)} \hat q^{\dag}_{j_1} \dots \hat q^{\dag}_{j_p}\ket{0}.
\label{S:Scars_partial}
\end{split}
\end{equation}
for the scarred eigenstate $\ket{\psi^{(l)}_{p,L/2}}$ ($\ket{\psi^{(r)}_{p,L/2}}$) defined on the half-chain $j=1,2\dots L/2$ ($j=L/2 +1, \dots L$). Since the sets of states $\mathcal{B}_l = \{\ket{\psi_{p,L/2}^{(l)}} \}_{p=0}^{L/2}$ and $\mathcal{B}_r = \{\ket{\psi_{p,L/2}^{(r)}} \}_{p=0}^{L/2}$ are orthonormal sets of states on the left and right half-chains, respectively, the expressions in Eq.~\eqref{S:Eq:SD_scars} are Schmidt decompositions of the scarred eigenstate $\ket{\psi_{n,L}}$ defined on the entire chain. The fundamental observation behind Eqs.~\eqref{S:Eq:SD_scars} and \eqref{S:Scars_partial} is that the sum over lattice sites in the definition of the quasi-particle creation operator \eqref{eq:quasi_particle_operator} can be decomposed into a part acting only on the left subsystem $j_p=1,2,\dots L/2$ and a part acting only on the right one $j_p=L/2+1,\dots L$, as discussed in Refs.~\cite{Schecter_2019,EE_scars_FH,Gotta_2023}. In doing this, one has to keep into account that for a subsystem size $L/2$, the quasiparticle creation operator can be applied at most $L/2$ times. This truncates the sum in Eq.~\eqref{S:Eq:SD_scars} to a number of terms equal to $\mathrm{min}(n,L/2) - \mathrm{max}(0,n-L/2)+1$, which gives the Schmidt rank of a scar eigenstate. The upper bound arises because the number of particles occupying one half of the system cannot exceed its length $L/2$, while the lower bound results from the fact that there must be a nonzero number of particles in each of the two halves of the system when $n>L/2$.

For the sake of concreteness, let us assume that $n\leq L/2$, and compute the reduced half-chain density matrix $\mbox{Tr}_{\{L/2+1,\dots,L\}}\Big[\ket{\psi_{n,L}}\bra{\psi_{n,L}}\Big]$. This yields the following diagonal ensemble on scarred eigenstates defined on half of the chain:
\begin{equation}
    \mbox{Tr}_{\{L/2+1,\dots,L\}}\Big[\ket{\psi_{n,L}}\bra{\psi_{n,L}}\Big]=\sum_{m=0}^n p_m \ket{\psi_{m,L/2}^{(l)}}\bra{\psi_{m,L/2}^{(l)}},
\label{S:reduced_half_chain_state}
\end{equation}
where:
\begin{equation}
    p_m = \frac{\binom{L/2}{m}\binom{L/2}{n-m}}{\binom{L}{n}}, \quad \sum_{m=0}^{n}p_m=1,
\end{equation}
are the eigenvalues of the reduced density matrix, properly normalized to one. The eigenvalue of the reduced density matrix is thus given by all the possible arrangements with $m$ quasi-particles in the left subsystem and $n-m$ quasi-particles in  the right one, normalized by the arrangements of $n$ quasi-particles into $L$ sites. Both Eq.~\eqref{S:reduced_half_chain_state} and \eqref{eq:diagonal_ensemble} are statistical mixtures of scarred eigenstates (in the former case the states are of course defined on an halved Hilbert space). The similarity between Eq.~\eqref{S:reduced_half_chain_state} and \eqref{eq:diagonal_ensemble} is made explicit by looking at the large-size scaling of the coefficients $p_m$. In the large size and quasiparticle number limit $L,m,n \to \infty$, with fixed ratios $m/L$ and $n/L$, we introduce the continuous variables $x=2m/L$ and $\rho=n/L$. The probability distribution $\{p_m\}_{m=0}^{n}$ can then be approximated by the continuous function $p(x;\rho)$:
\begin{equation}
    p(x;\rho) = \sqrt{\frac{2}{\pi L \rho(1-\rho)}} e^{-\frac{L}{2\rho(1-\rho)}(x-\rho)^2}.
\end{equation}
This probability preserves the correct normalization of the eigenvalues
\begin{equation}
    1=\sum_{m=0}^n p_m \approx \frac{L}{2}\int_0^{2\rho} dx\, p(x;\rho).
\end{equation}
We therefore see that $p(x;\rho)$ features gaussian concentration as a function of $x$ around the value $x^{\ast}=\rho$, similarly to the concentration of the coefficients $|c_n|^2$ around $y^{\ast}$ in Eq.~\eqref{S:y_star_saddle}. The asymptotic scaling as a function of $L$ of all the Rényi and entanglement entropies can thus be evaluated by saddle point analysis similarly to what we did for the mixedness in Eq.~\eqref{S:Eq:asympt_fid}. For the entanglement entropy, for instance, one obtains \cite{Schecter_2019,EE_scars_FH}  
\begin{equation}
S\left(\mbox{Tr}_{\{L/2+1,\dots,L\}}\Bigl[\ket{\psi_n}\bra{\psi_n}\Bigr] \right)=-\sum_{m=0}^{n}p_m \log p_m = \log\left[\sqrt{\pi \rho(1-\rho)L}\right].
\end{equation}
This asymptotic behavior is analogous to that in Eq.~\eqref{S:Eq:small_r_estimate} for the mixedness obtained in the presence of resetting.

\subsection{Time dependence of the $2$-Rényi-entropy dynamics after a quench from a coherent state}
\label{app:time_dependence_Renyie}
We compute in this subsection the transient dynamics of the $2$-Rényi entropy prior to reaching the steady state. To this end, we start by introducing a coherent state defined over a system of $L$ sites as:
\begin{equation}
    \ket{\alpha(t)} = \frac{1}{(1+|\alpha|^2)^{L/2}}\prod_{j=1}^L (1+\alpha e^{ikj+i\omega t}\hat q^{\dag}_j)\ket{0};
\end{equation}
then, we write the full density matrix of the system in presence of a finite resetting rate from the renewal equation:
\begin{equation}
    \hat \rho_r(t) = e^{-rt}\ket{\alpha(t)}\bra{\alpha(t)} + r\int_0^t d\tau\, e^{-r\tau} \ket{\alpha(\tau)}\bra{\alpha(\tau)}.
\end{equation}
By taking the square of the density matrix $\hat\rho_{r}(t)$, one obtains:
\begin{equation}
\begin{split}
    &\hat \rho_{r}^2(t) = e^{-2rt}\ket{\alpha(t)}\bra{\alpha(t)}+r^2 \int_0^t d\tau_1 \int_0^t d\tau_2 e^{-r(\tau_1+\tau_2)}\ket{\alpha(\tau_1)}\braket{\alpha(\tau_1)|\alpha(\tau_2)}\bra{\alpha(\tau_2)}+\\
    &+re^{-rt} \int_0^t d\tau e^{-r\tau} \ket{\alpha(t)}\braket{\alpha(t)|\alpha(\tau)}\bra{\alpha(\tau)}+re^{-rt} \int_0^t d\tau e^{-r\tau} \ket{\alpha(\tau)}\braket{\alpha(\tau)|\alpha(t)}\bra{\alpha(t)}.
    \end{split}
\end{equation}
After some manipulations, the result of taking the trace of the above expression and then its logarithm to obtain the $2$-Rényi entropy reads:
\begin{equation}
\begin{split}
    &S_{2}(\hat\rho_r(t))=-\log \left(\mbox{Tr}[\hat \rho_{r}^2(t)] \right)=\\
    = -\log &\Biggl[e^{-2rt}+\frac{2re^{-rt}}{(1+|\alpha|^2)^{2L}}\sum_{l=0}^{L}\sum_{m=0}^{L}\binom{L}{l}\binom{L}{m}|\alpha|^{2(l+m)} e^{i\omega(m-l)t}\frac{1-e^{-rt+i\omega(l-m)t}}{r-i\omega(l-m)}+\\
    &+\frac{r^2}{(1+|\alpha|^2)^{2L}}\sum_{l=0}^{L}\sum_{m=0}^{L}\binom{L}{l}\binom{L}{m}|\alpha|^{2(l+m)} \left|\frac{1-e^{-rt+i\omega(l-m)t}}{r-i\omega(l-m)} \right|^2 \Biggr].
    \end{split}
\label{eq:t_dependent_S2}
\end{equation}
The numerical evaluation of this expression lead to the results presented in Fig.~\ref{Fig:4} of the main text for the dynamics of the Rényi-$2$ entropy as a function of time.

\section{Local equivalence to a quantum scarred eigenstate}\label{S:local_equivalence}

In this section, we show that the state \eqref{eq:diagonal_ensemble} describing the weak reset limit is locally equivalent in the thermodynamic limit to a scarred state with index $n^{\ast}=y^{\ast}(\alpha)L$. To do this, we consider local operators $\hat O$ and compute the expectation value over the state \eqref{Eq:diag_obs}
\begin{equation}
    \mbox{Tr}[\hat O \hat \rho_{0^+}(+\infty;\alpha)] = \sum_{n=0}^L |c_n|^2 \bra{\psi_n}\hat O \ket{\psi_n}\approx \frac{L}{(1+|\alpha|^2)^L}\int_0^1 dy\, \frac{e^{-L f(y)}}{\sqrt{2\pi L y(1-y)}} \bra{\psi_{\lfloor Ly\rfloor}}\hat O\ket{\psi_{\lfloor Ly \rfloor}},
\end{equation}
with the function $f(y)$ in Eq.~\eqref{eq:saddle_point_eq}. In the second equality we transformed in the thermodynamic limit $n,L\to \infty$, with fixed ratio $n/L$, the discrete sum over the index $n$ to an integral over the coordinate $y=n/L$, as similarly done for Eqs.~\eqref{Eq:diag_2}-\eqref{S:Eq:asympt_fid}. The integral can then be evaluated by saddle-point and it gives
\begin{equation}
  \lim_{L\to\infty}\lim_{r\to 0^+}  \mbox{Tr}[\hat O \hat \rho_{0^+}(+\infty;\alpha)] =\bra{\psi_{\lfloor Ly^{\ast}\rfloor}}\hat O\ket{\psi_{\lfloor Ly^{\ast} \rfloor}}.
\label{S:Eq:local_equivalence}
\end{equation}
This equality establishes the equivalent between the diagonal steady state obtained from weak resetting and single pure scarred eigenstate with index $n^{\ast}=L y^{\ast}$ given by Eq.~\eqref{S:y_star_saddle}. The equality in Eq.~\eqref{S:Eq:local_equivalence} crucially applies for local observables only, this is the reason why we refer to it as local equivalence. For nonlocal observables, indeed, the expectation value $\bra{\psi_{\lfloor Ly\rfloor}}\hat O\ket{\psi_{\lfloor Ly \rfloor}}$ can exponentially depend on the system size $L$. This can lead to a shift of the saddle point compared to the value $y^{\ast}$.

To exemplify this aspect we detail the calculation of the fidelity presented in Eqs.~\eqref{eq:left_fidelity}-\eqref{eq:right_fidelity} of the main text in the thermodynamic limit. For the fidelity one, indeed, needs to consider the nonlocal operator $\ket{\alpha}\bra{\alpha}$ into Eq.~\eqref{S:Eq:local_equivalence}. The left-hand side of the equation reads
\begin{align}
 \mbox{Tr}[\ket{\alpha}\bra{\alpha}\hat{\rho}_0(+\infty)] &= \sum_n |c_n|^2 |\braket{\alpha=1 | \psi_n}|^2 = \sum_n |c_n|^4 = \frac{1}{(1+|\alpha|^2)^{2L}} \sum_n \binom{L}{n}^2 |\alpha|^{4n}\Biggl|_{\alpha=1}\approx\\
     &\approx \frac{L}{2^{2L}}\int_0^1 dy \frac{e^{-2Lf(y)}}{2\pi L y(1-y)}\approx \frac{1}{\sqrt{\pi L}},  
\label{S:E1}
\end{align}     
with a calculation analogous to the one done in Eqs.~\eqref{Eq:diag_2}-\eqref{S:Eq:asympt_fid} for the second Rényi entropy. On the other hand, the squared overlap between the target scar state with density $n=L/2$ is given by:
\begin{equation}  \Bigl|\braket{\alpha=1 | \psi_{L/2}}\Bigr|^2 = \frac{1}{2^L}\binom{L}{L/2}\approx \sqrt{\frac{2}{\pi L}},
\label{SE:2}
\end{equation}
where in the second equality we used Eq.~\eqref{S:coefficients} for the overlaps between the state $\ket{\alpha}$ and scarred eigenstates. We therefore see that Eq.~\eqref{S:E1} differs from \eqref{SE:2} by a factor $\sqrt{2}$. The latter arises because in Eq.~\eqref{S:E1} the overlap $|\braket{\alpha=1|\psi_n}|^{2}$ depends exponentially on the system size and leads to a doubling of the function $f(x)$ determining the saddle point. This further confirms that Eq.~\eqref{S:Eq:local_equivalence} applies to local observables only.

\end{document}